\documentstyle[graphicx,supertabular]{aa}

\newcommand{\be}{\begin{equation}}
\newcommand{\ee}{\end{equation}}

\begin{document}

\authorrunning{Mauro Dadina}
\titlerunning{Average X-ray properties of Seyferts observed by $BeppoSAX$}

\title{Seyfert galaxies in the local Universe (z$\leq$ 0.1): the average X-ray spectrum as seen by $BeppoSAX$}

\author{Mauro Dadina
\inst{1,}
\inst{2}}

\institute{$^{1}$INAF/IASF-Bo, via Gobetti 101, 40129 Bologna, Italy\\
$^2$Dipartimento di Astronomia dell'Universit\`a degli Studi di Bologna, 
via Ranzani 1, 40127 Bologna, Italy}

%\email{mauro.dadina@iasfbo.inaf.it}

\date{Received date/ Accepted date}

\abstract{

The $BeppoSAX$ archive is currently the largest reservoir of high 
sensitivity simultaneous soft and hard-X ray data of Seyfert galaxies. 
From this database all the Seyfert galaxies (105 objects of which 43 are type I
and 62 are type II) with redshift lower than 0.1 have 
been selected and analyzed in a homogeneous way (Dadina 2007). 
Taking advantage of the broad-band coverage of the $BeppoSAX$ MECS and PDS 
instruments ($\sim$2-100 keV), the X-ray data so collected allow 
to infer the average spectral properties of nearby 
Seyfert galaxies included in the original sample and, most notably: the 
photon index ($\Gamma$$\sim$1.8), the high-energy cut-off (Ec$\sim$290 keV), 
and the relative amount of reflection (R$\sim$1.0).
The data collected have been also used to test some basic assumptions of 
the unified scheme for the active galactic nuclei. The 
distributions of 
the isotropic indicators used here (photon index, relative amount of 
reflection, high-energy cut-off and narrow FeK$\alpha$ energy centroid) are 
similar in type I and type II objects while the absorbing column and the 
iron line equivalent width significantly differ between the two classes of active galactic nuclei with type II objects displaying larger columns (N$_{H}$$\sim$3.7$\times$10$^{22}$ and 6.1$\times$10$^{23}$ cm$^{-2}$ for type I and II objects respectively) and equivalent width (EW$\sim$220 and 690 eV for type I and II 
sources respectively).
Confirming previous results, the narrow FeK$\alpha$ line is consistent, in 
Seyfert 2, with being produced in the same matter responsible for the observed 
obscuration. These results, thus, support the basic picture of the unified 
model. Moreover, the presence of a 
X-ray Baldwin effect in Seyfert 1 has been here measured  using the 20-100 keV 
luminosity (EW$\propto$L(20-100)$^{-0.22\pm0.05}$). Finally, the possible presence of a correlation  
between the photon index and the amount of reflection is confirmed thus 
indicating thermal Comptonization as the most likely origin of the high energy 
emission for the active galactic nuclei included in the original sample.


}

\maketitle

\section{Introduction}

The high energy emission from active galactic nuclei (AGN) is thought to come 
from the innermost  regions of accretting systems that are centered around 
super-massive black-holes (SMBH). For this reason, X-rays are expected to 
be tracers of the physical conditions experienced by matter before 
disappearing into SMBH. Moreover, thanks to their high penetrating power, 
energetic photons, escaping from the nuclear zones, test the matter located 
between their source and the observer. Thus, they offer powerful diagnostics 
to understand the geometry and the physical conditions of the matter 
surrounding the SMBH.

The broad-band of $BeppoSAX$ offered for the very first time the opportunity 
to measure with a remarkable sensitivity, the spectral shape of AGN in the 
0.1-200 keV range. This potential has been previously exploited to study in 
details a number of sample selected in different manners (see for example 
Maiolino et al. 1998; Perola et al. 2002). These studies were fundamental in
making 
important steps forward in the comprehension of the emitting mechanism at work 
in the production of X-rays (Perola et al. 2002) and to partially unveil the 
geometry of the cold matter surroundings the central SMBH 
(Maiolino et al. 1998; Bassani et al. 1999; Risaliti et al. 1999).

The $BeppoSAX$ database full potential, however, was never exploited 
before. In a previous paper, the entire catalog of the Seyfert galaxies
at z$\leq$0.1 contained in the $BeppoSAX$ archive has been presented 
(Dadina 2007). This sample was selected starting 
from the catalog of Seyfert galaxies contained in the V{\'e}ron-Cetty \& V{\'e}ron (2006) sample of AGN and contains 13 radio-loud objects and 7 narrow-line 
Seyfert 1. 

The spectral analysis was performed in 
the 2-100 keV band whenever possible and the data were fit with a set of 
template models to obtain a homogeneous dataset. Here the X-ray data so 
collected are statistically inspected
in order to infer what are the average characteristics of the nearby Seyfert 
galaxies contained in this sample in the 2-100 keV band. Finally, 
present dataset is used to perform simple tests on the 
unified scheme (UM) for the AGN (Antonucci 1993) and 
on the emission mechanism acting in the core of the Seyfert galaxies. 
More detailed analysis on this latter 
topic will be presented in another paper (Petrucci, Dadina \& Landi, 
submitted) where detailed thermal Comptonization models (Poutanen 
\& Svensson 1996, Haardt \& Maraschi 1993 ) will be used to fit the 
BeppoSAX data with the main scope to study the dependence of the 
spectral properties in the ``two phase'' scenario (Haardt 1991, Haardt \& 
Maraschi 1993) assuming different geometries of the corona.

\section{Mean X-ray Properties}

Scope of this section is to determine and study the mean X-ray spectral 
properties of the sample and to use these values to test the UM model for AGN 
(Antonucci 1993). The key parameters are the ones that 
describe the continuum and the absorption properties. In the framework
of the UM models for AGN, the continuum shape is expected to
be independent from the orientation angle under which the 
source is observed. Thus, no difference should be measured in the parameters 
describing the continuum between type I and 
II objects. On the contrary, the absorbing column intervening in Seyfert 2
should be the principal discriminator between the two classes of objects. Thus
measuring the mean X-ray properties means to test the basic assumptions
of the UM.

\subsection{Methods}

The origins of the X-ray photons from AGN are thought to be due to 
Comptonization of optical-UV radiation, coming from the accretion disk 
and Comptonized by the e$^{-}$ in the hot corona that sandwiches the disk (Haardt \& Maraschi 1991; Haardt 1993, Poutanen \& Svensson 1996, Czerny  et al. 2003). The mechanism is assumed to 
be, at least at the zero-th order, very similar in each Seyfert galaxy. 
Under this hypothesis, the differences 
between the X-ray spectra of different objects are supposed to be mainly due 
to two kind of factors: i) the time-dependent state of the emitting 
source; ii) the intervening matter that imprints on the emerging spectrum the 
features typical of its physical state. In such a scenario, the observations 
of 
many sources can be regarded as the long-term monitoring of a single source. 
On the other hand, it is also true that the contrary has some comparison in the
 literature: e.g. time sparse observations of a single source in different states can resemble observations of objects with completely different spectral properties. This is the case, for example, of the narrow line Seyfert 1 NGC 4051 
that displayed variations in flux/luminosity by a factor of $\sim$100 
(Guainazzi et al 1998) associated to strong variations of the spectral shape 
($\Gamma$$\sim$0.5-2.4; Guainazzi et al. 1998; Turner et al. 1999; Ponti et al.
 2006,  but see also Crenshaw \& Kraemer (2007) for a different 
interpretation of this spectral behavior in terms of variable ionized 
absorption). To calculate the mean X-ray 
properties of the sources included in this work, thus, I treated the multiple 
observations of single sources separately: i.e. I assumed different 
observations of the same source as if they were observations of different 
sources.

This method, in principle, is expected to work properly for all those 
quantities which are supposed to vary in accordance with the state of activity
of the central nucleus. For example, the 
photon index $\Gamma$ is known to vary accordingly with
the AGN flux state (Lee et al. 2000;  Shih, Iwasawa \& Fabian 2002, 
Ponti et al. 2006) and  the high-energy
cut-off (Ec) is linked with the temperature of the corona and thus
expected to be variable (Haardt, Maraschi \& Ghisellini 1997). On the contrary 
this method is not expected to work 
properly when constant components are considered. This could be the case, 
for instance, of the cold absorption assumed to be due to the putative
dusty torus (Antonucci 1993). For this component, thus, the average value
recorded for each source should be used. Nonetheless, also these components 
were observed to vary in a number of objects (see for example the case of 
NGC 4151, De Rosa et al. 2006 or Risaliti et al. 2002). Moreover, the 
constancy of the properties of the cold absorption is predicted under the 
hypothesis of a continuous distribution of the matter that forms the torus. On the contrary, if the torus is formed by blobs/clouds 
(Elitzur \& Shlosman 2006), variability in the measurements of absorbing column
is naturally expected. 
For both these observational and theoretical reasons also the N$_{H}$ measured 
in each single observation were treated separately.

In between these two cases, are the properties of the emission
FeK$\alpha$ line. $BeppoSAX$ had a too low sensitivity to detect the 
relativistically
broadened component of this feature in a large number of sources. Thus 
it turned-out that only the narrow line has been detected in the vast 
majority of the objects included  in the original catalog. 
This component is supposed to originate 
far from the SMBH, at least at $\sim$1000 Schwarzschild radii 
(Mattson \& Weaver 2004), i.e. very likely at the inner edge of the torus 
(Nandra 2006). At these distances from the SMBH, the relativistic 
effects are negligible.
Thus, in principle, the parameters that describe the line should be 
regarded as stable. Nonetheless also the properties of the narrow
FeK$\alpha$ line were observed to vary with  its line energy
centroid changes between $\sim$6.4 keV (neutral Iron) and $\sim$6.9 keV 
(H-like Iron) showing different 
ionization states. Moreover, the equivalent width (EW) of 
this component is not only function of the intensity of the line itself, but
it is also directly linked with the underlying variable continuum emitted 
in the regions close to the SMBH. 
This makes EW a variable quantity. The parameters of the emission
FeK$\alpha$ line were, thus, treated as variable ones in order to test its 
origin, and not averaged when  objects were observed more than once.

Finally, it is worth noting here that in a number of cases, it was necessary 
to deal with censored data (see table 1). 
To manage these data properly, the ASURV software 
(Feigelson \& Nelson 1985; Isobe et al. 1986) has been used. 
In particular, to establish if the distributions of parameters of type I 
and type II objects were drawn from different parent populations, the   
Peto \& Prentice generalized Wilcoxon test (Feigelson \& Nelson 1985) has been 
used while to calculate the mean values considering also the censored data
the Kaplan-Meyer estimator has been used. To establish the presence of 
correlations between different quantities, both the Spearman's $\rho$ and the 
generalized Kendall's $\tau$ methods were applied. The linear regressions were 
calculated using the Bukley-James and Schmitt`s methods. In the following, 
two quantities are considered as drawn from different parent population
when the probability of false rejection of the null 
hypothesis (same parent population) is P$_{null}$$\leq$1\%.
Similarly, one accepts that there is a correlation between two given quantities 
when the probability of absence of correlation remains  smaller than 1\%.

\vspace{0.5cm}

\small
\tablecaption{ General characteristics of the data analyzed in this work. 
The number of detections and censored data are reported for the interesting
parameters for the whole sample of objects (columns 2 and 3), for the 
Seyfert 1 galaxies (columns 4 and 5), and for the Seyfert 2 objects (column 6 
and 7). The 90\% confidence interval limits were used for censored data and
the detected values were defined if determined with a 99\% confidence level (Dadina 2007)}
\begin{supertabular}{ l  c c  c c  c c }
\hline
\hline
& & & & & &\\

\multicolumn{1}{c}{Parameter}&\multicolumn{2}{ c }{Tot. Sample}&\multicolumn{2}{ c }{Seyfert 1}&\multicolumn{2}{c}{Seyfert 2}\\
& & & & & & \\
& Det.&  Cens. &  Det. & Cens. & Det. & Cens.\\
& & & & & &\\
\hline
& & & & & & \\
N$_{H}$ & 83 & 80 & 31 & 53 & 52 & 27 \\
& & & & & & \\
EW& 129 & 7 & 66 & 4 & 63 & 3 \\
& & & & & & \\
R& 68 & 18 & 46 & 9 & 22 &9 \\
& & & & & & \\
Ec & 33& 51& 27& 26& 6 & 25\\
 & & & & & & \\

\hline\hline

\end{supertabular}

\normalsize

\vspace{0.3cm}

\subsection{The X-ray continuum and the cold absorption}

\begin{figure*}
\centering
\includegraphics[width=5cm,height=3cm,angle=0]{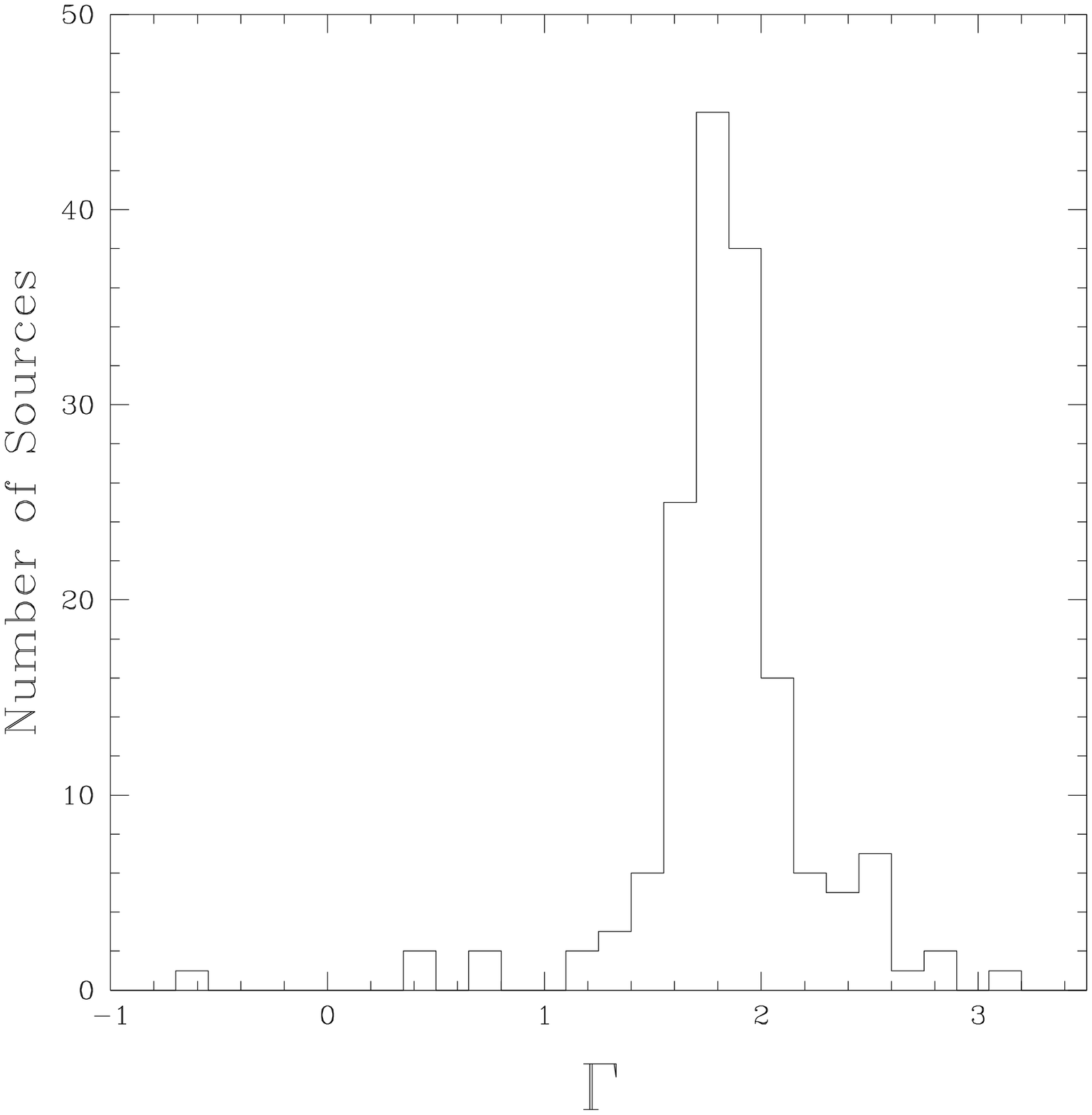}\includegraphics[width=5cm,height=3cm,angle=0]{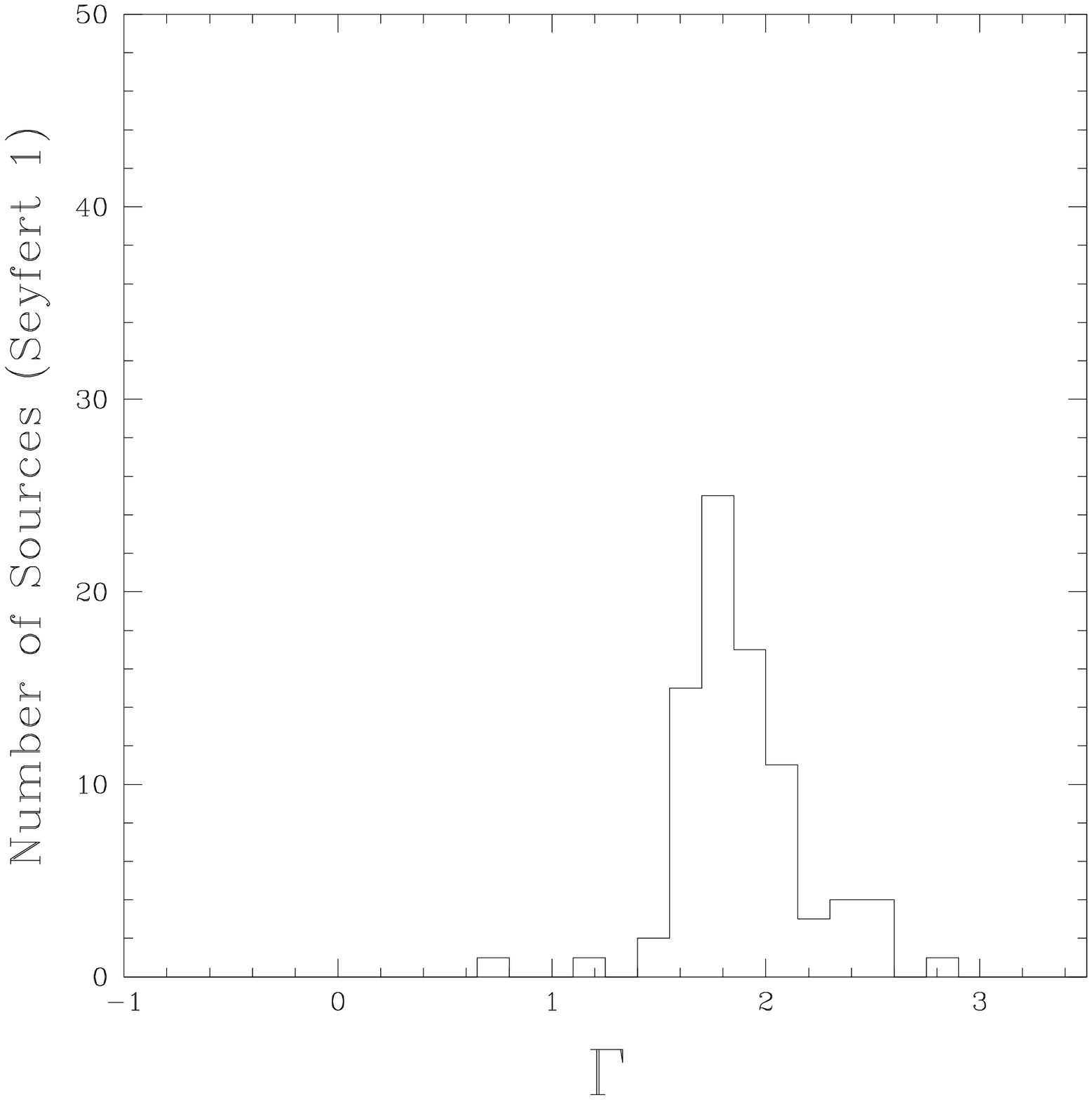}\includegraphics[width=5cm,height=3cm,angle=0]{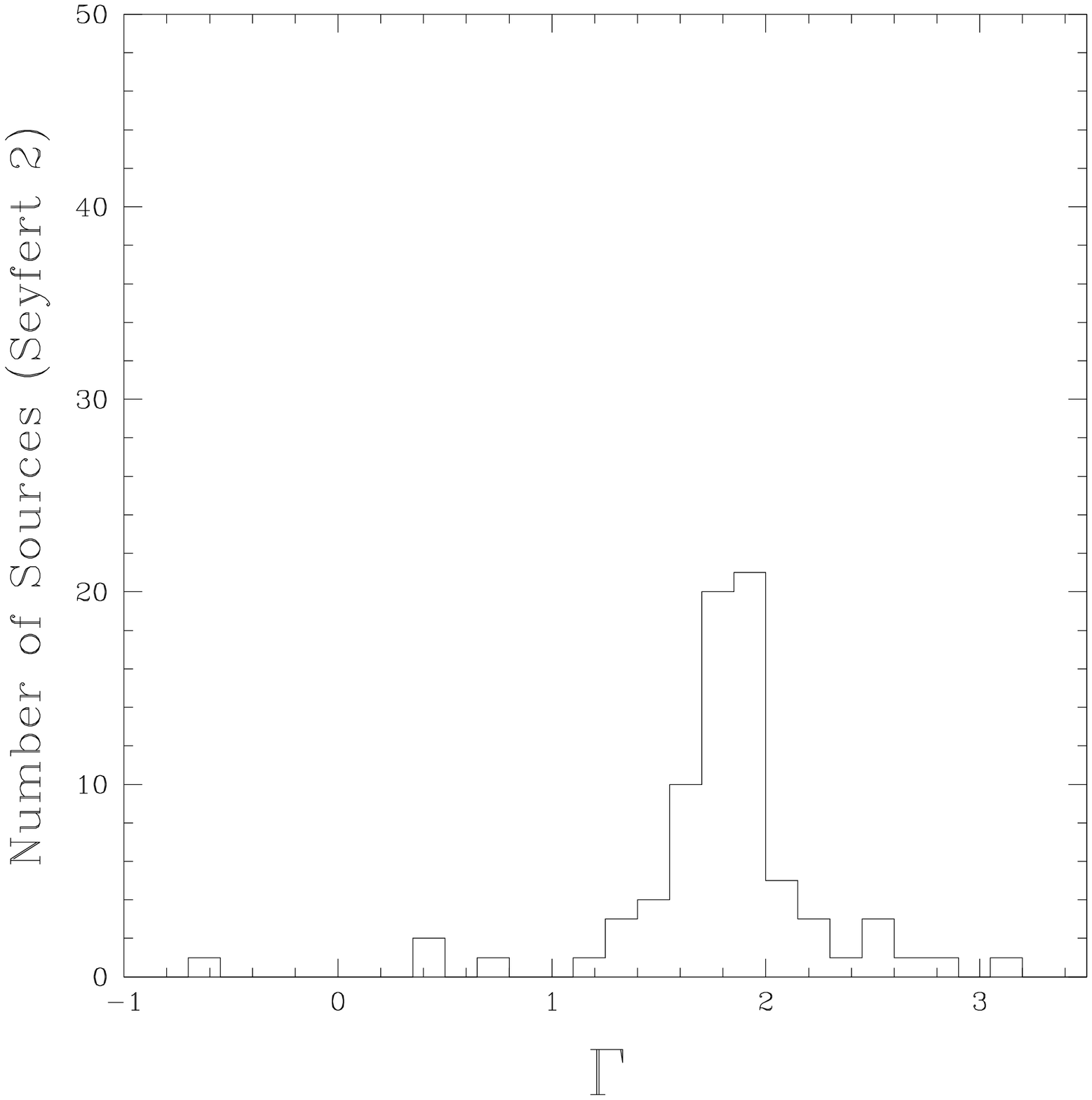}

\includegraphics[width=5cm,height=3cm,angle=0]{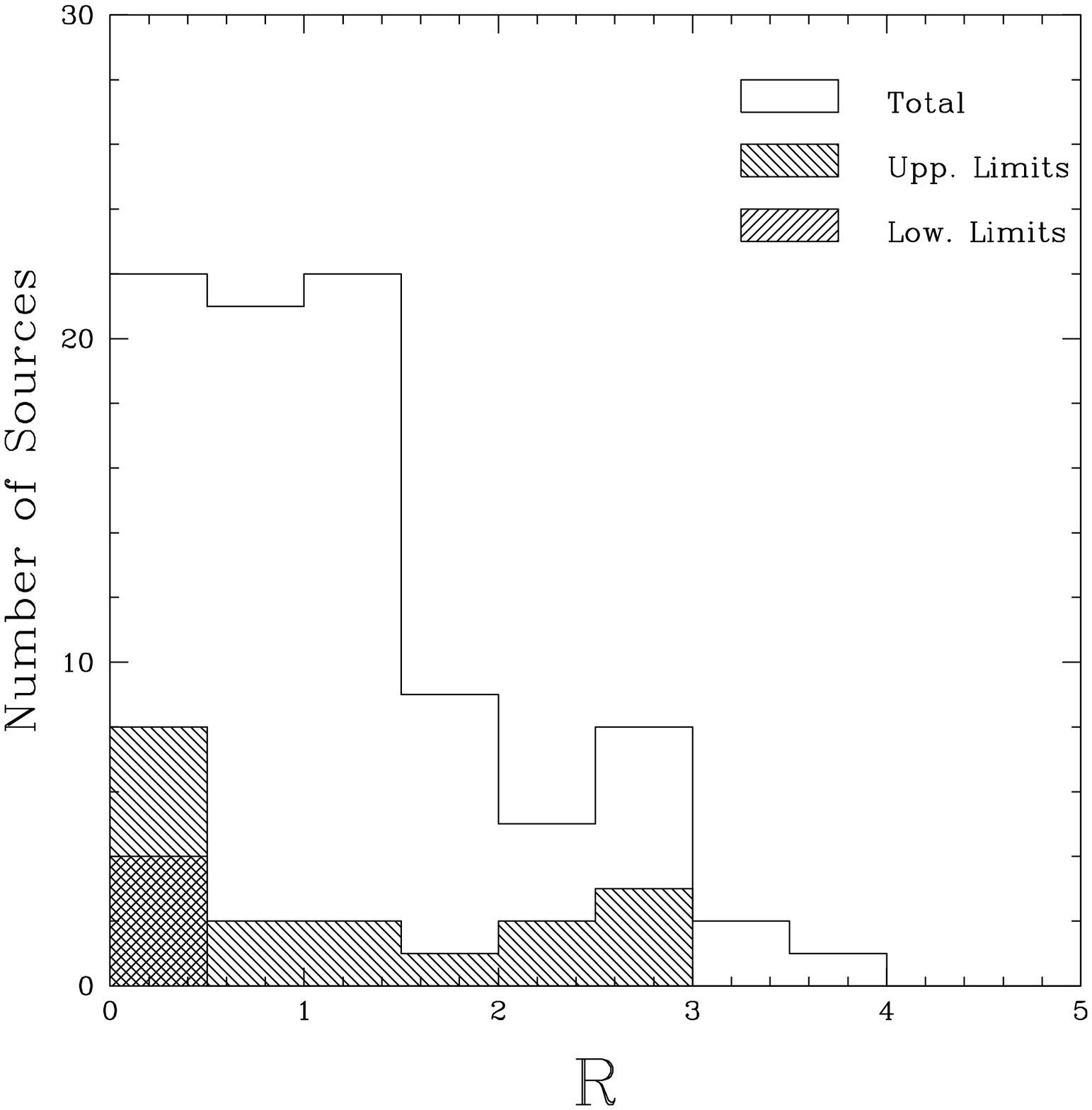}\includegraphics[width=5cm,height=3cm,angle=0]{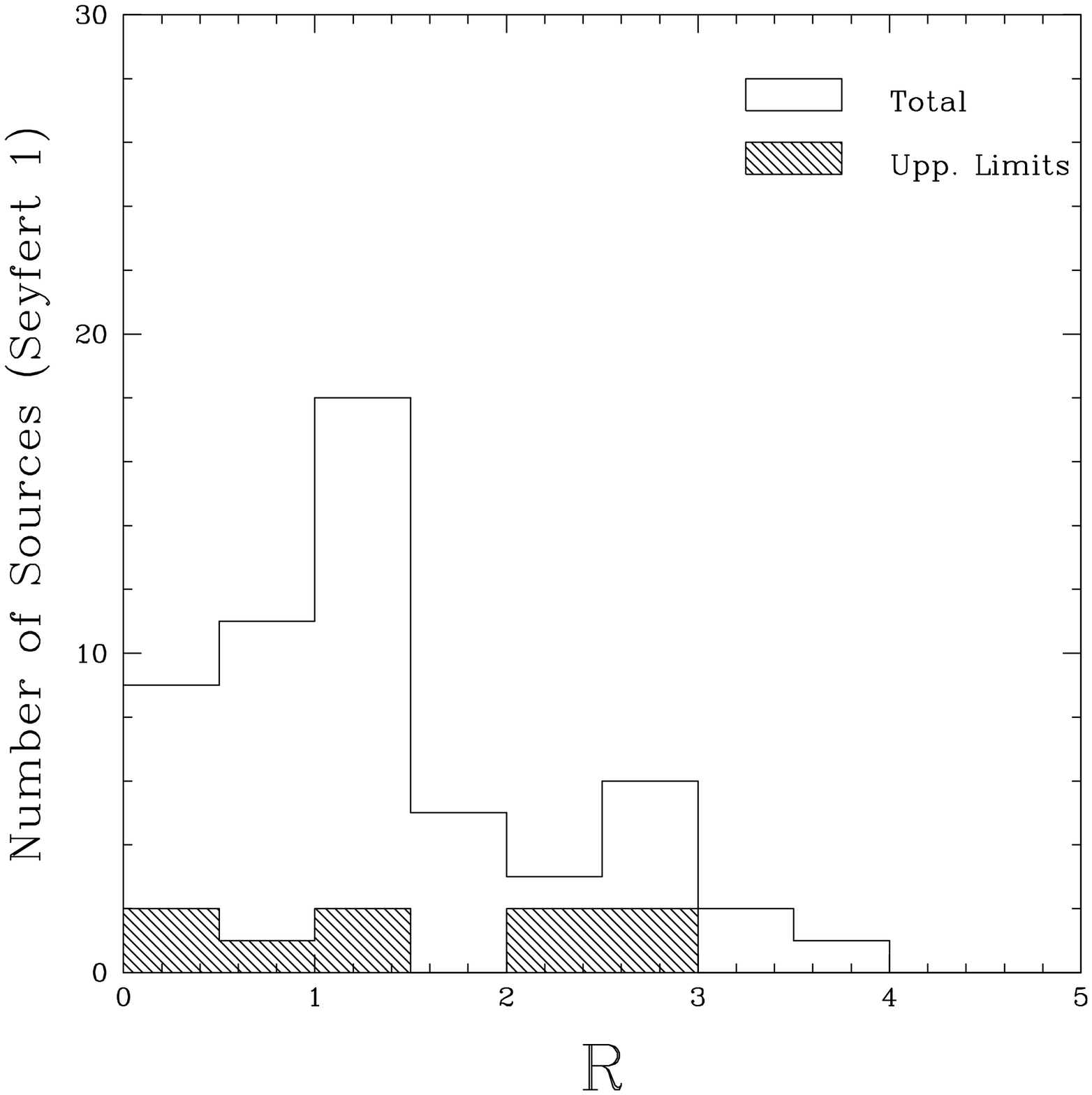}\includegraphics[width=5cm,height=3cm,angle=0]{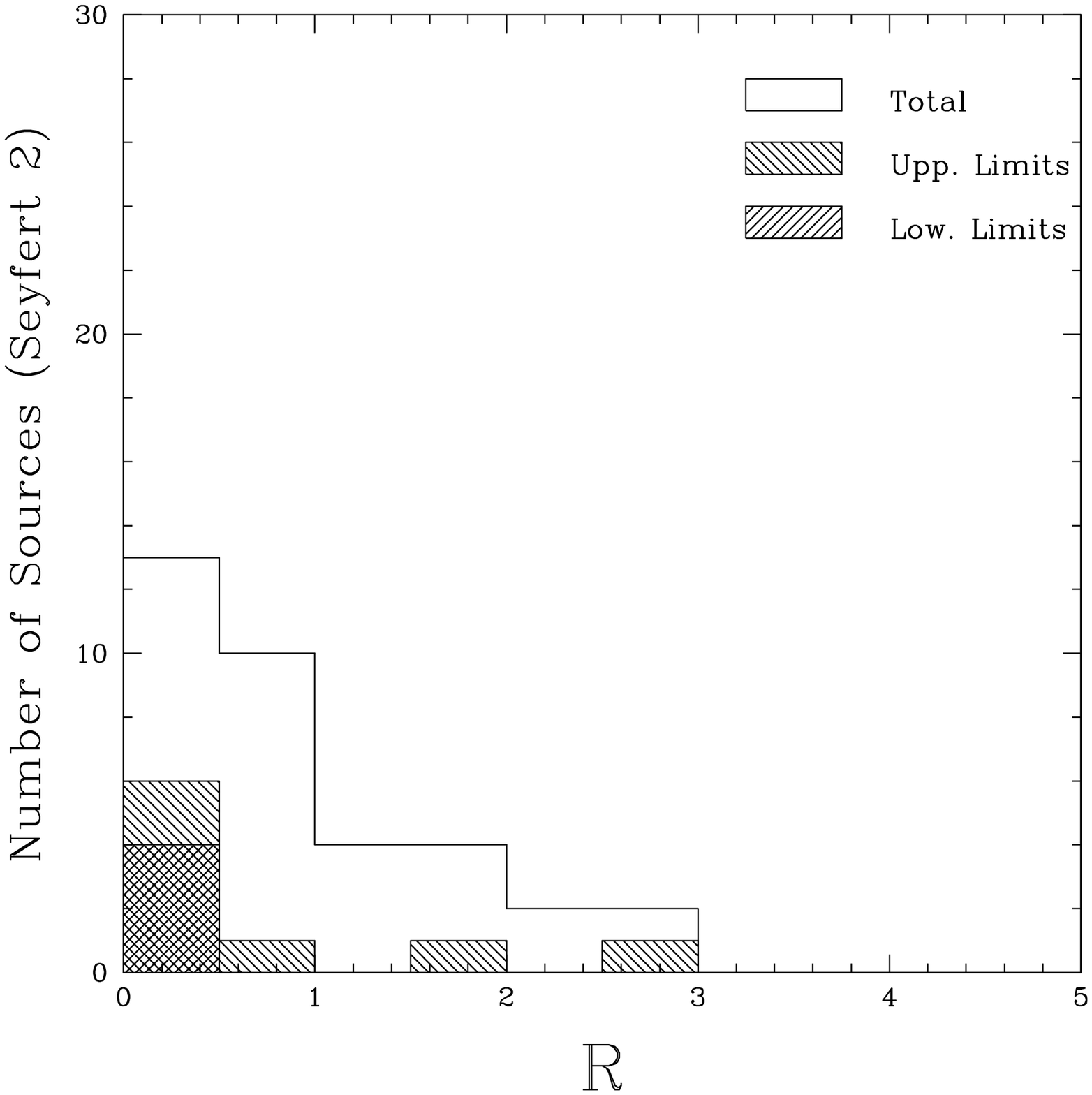}

\includegraphics[width=5cm,height=3cm,angle=0]{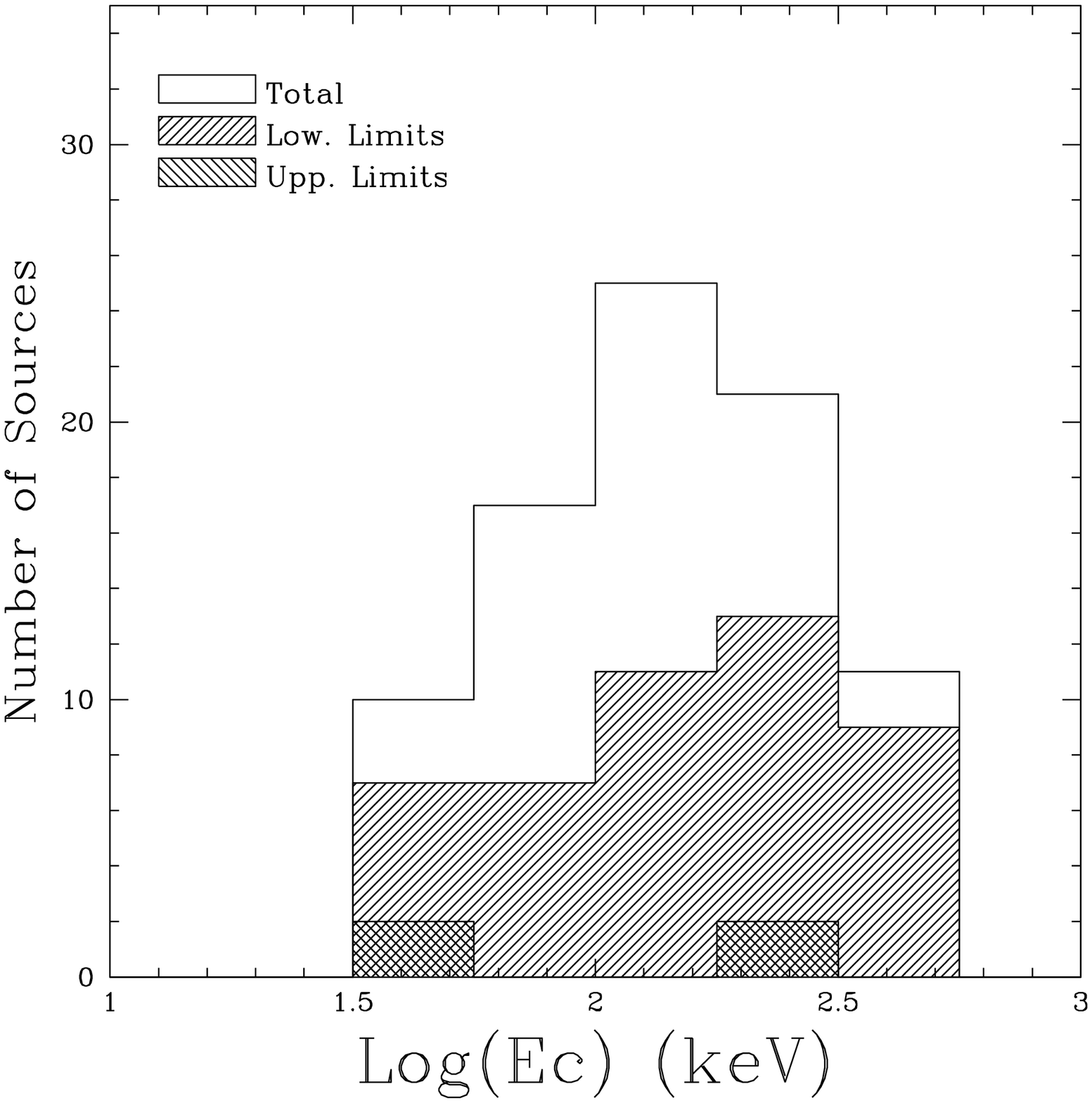}\includegraphics[width=5cm,height=3cm,angle=0]{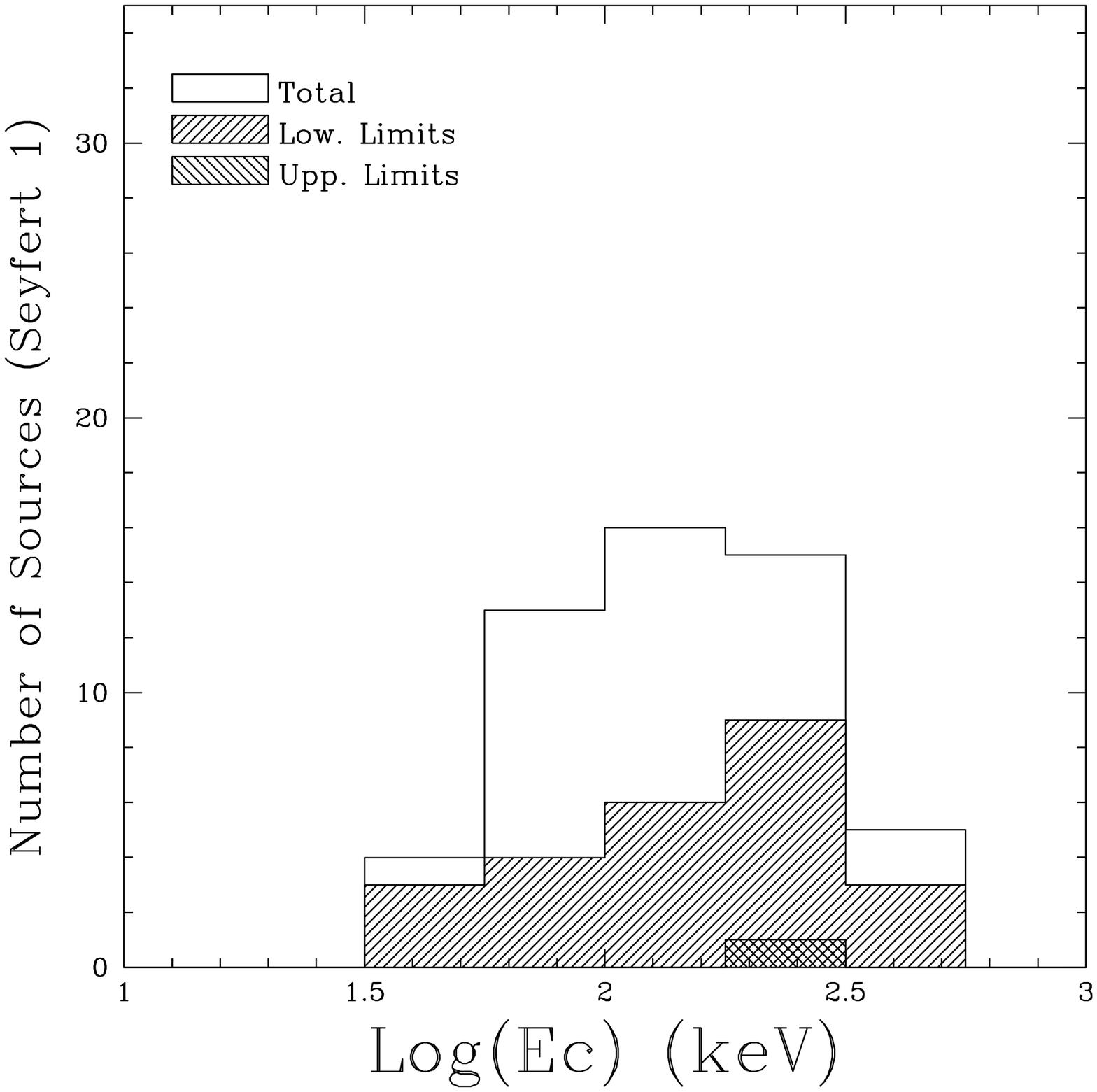}\includegraphics[width=5cm,height=3cm,angle=0]{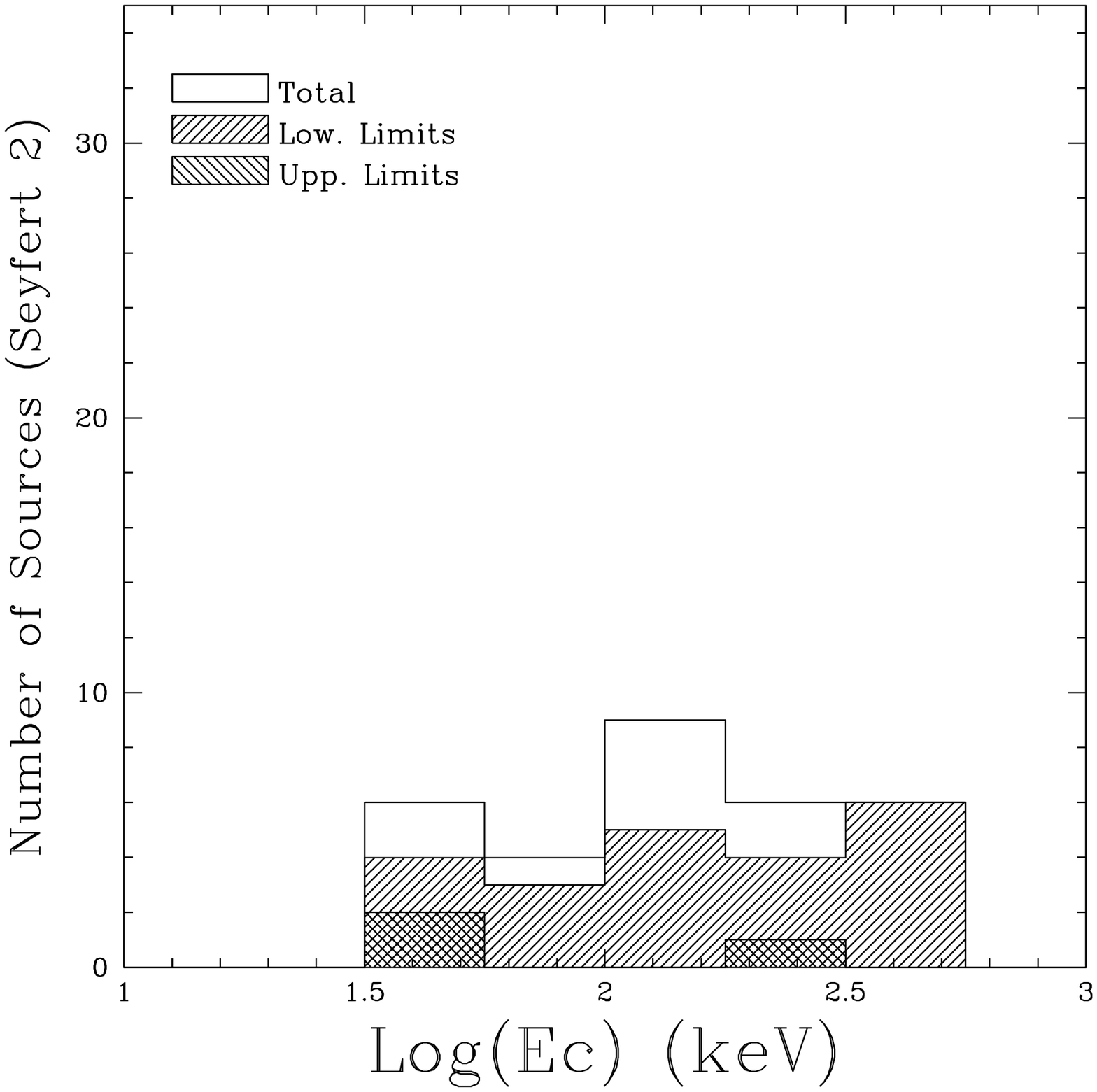}

\includegraphics[width=5.cm,height=3.cm,angle=0]{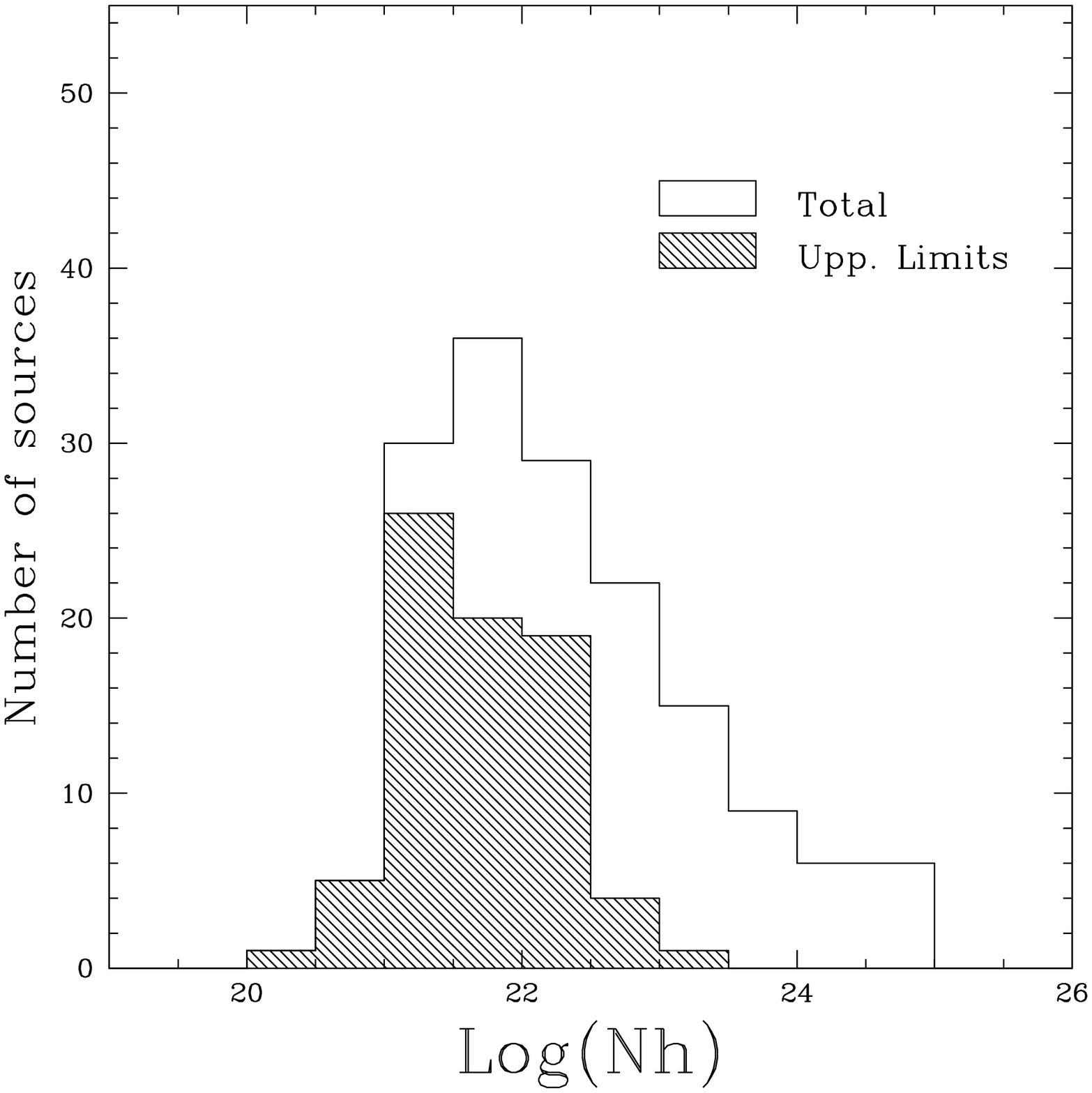} 
\includegraphics[width=5.cm,height=3.cm,angle=0]{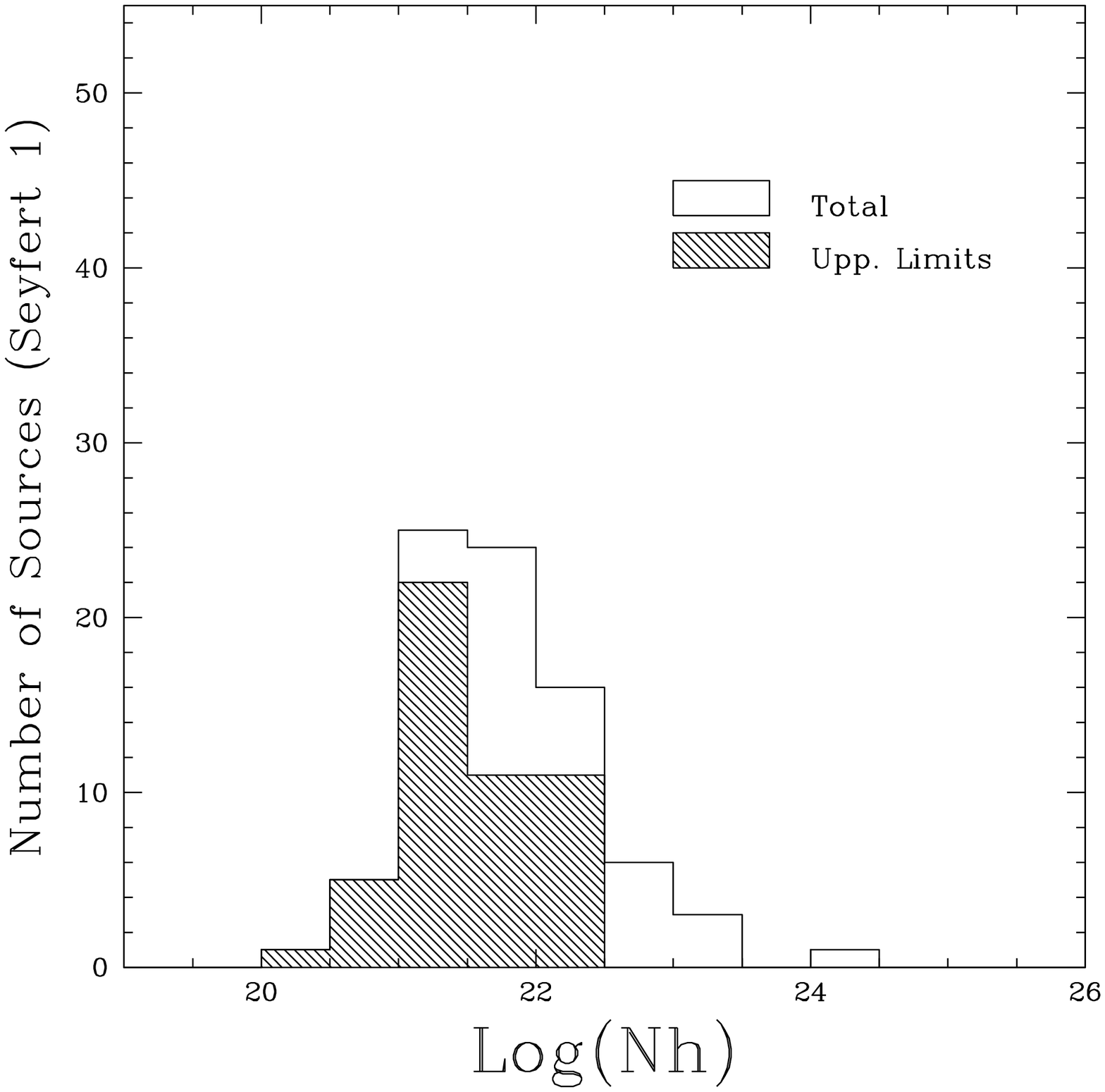}
\includegraphics[width=5.cm,height=3.cm,angle=0]{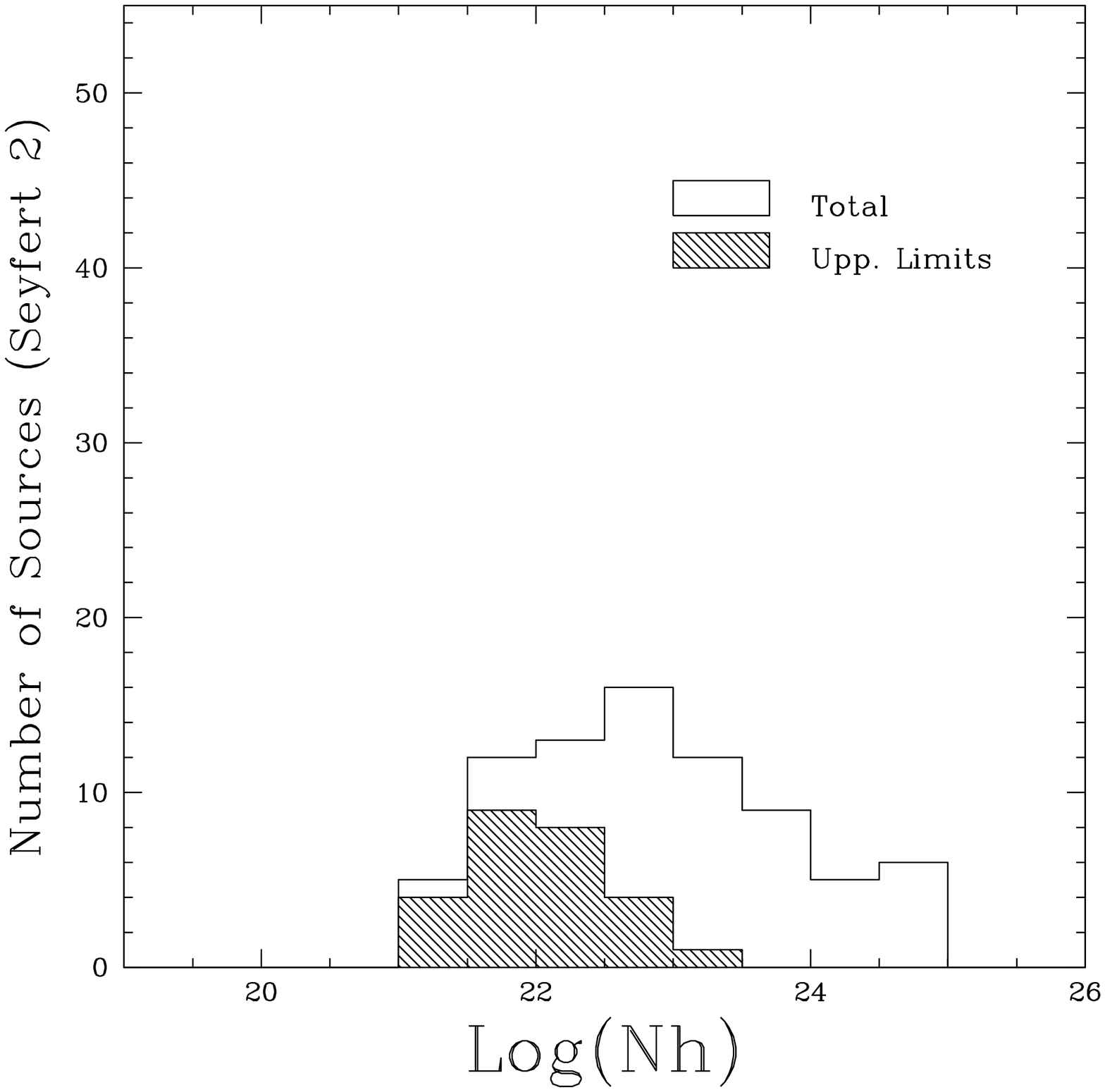}

\caption{Photon index $\Gamma$ ({\it first row}), R ({\it second row}), Ec ({\it third row}), and N$_{H}$ ({\it fourth row}) in units of cm$^{-2}$ distributions for the whole dataset ({\it left column}), for type I objects ({\it center column}), and for type II objects ({\it right column}).}

\end{figure*}

\vspace{-0.1cm}
The X-ray continua of the sources have been modeled using a cut-off power-law, 
that describes the primary emission  from the hot corona plus a 
reflection component ($PEXRAV$ model in Xspec, Magdziarz \& Zdziarski 1995) 
to account for the contribution expected to be due to the disk.
It is worth noting, however, 
that an additional reflection component could rise from the torus (Ghisellini, 
Haardt \& Matt 1994) and that the disentangling between the two reflection 
component is impossible with the quality of the available data. 
Wherever its origin, the reflection has been assumed to be due to cold matter.
The interesting parameters 
are the photon index ($\Gamma$), the relative amount of reflection 
(R), and  the high-energy cut-off (Ec).

In table 2 the results for the whole set of observations and for the 
two classes of Seyferts are reported as well as the probability that 
Seyfert 1 and 2 are drawn from  the same parent populations. The histograms 
of the distributions of the interesting parameters for 
the entire sample of objects (first column) and for type I (second column) and type II (third column) are reported in figure 1. 

\vspace{0.5cm}
\small
\tablecaption{Mean spectral properties. Col. I: Spectral parameter; Col. II: Seyfert 1 mean value; Col.III: Seyfert 2 mean value; Col. IV: Probability that Seyfert 1 and Seyfert 2 are drawn from the  same parent populations.}
\begin{supertabular}{ l c c c c }
\hline
\hline
& & & & \\

Parameter & Tot. & Seyfert 1 & Seyfert 2 & P$_{null}$ \\

& & & & \\

\hline

& & & & \\

$\Gamma$ & 1.84$\pm$0.03  & 1.89$\pm$0.03 & 1.80$\pm$0.05 & 90\% \\

& & & & \\

R & 1.01$\pm$0.09 & 1.23$\pm$0.11& 0.87$\pm$0.14 & 5\% \\

& & & & \\

Ec$^{\dagger}$& 287$\pm$24& 230$\pm$22& 376$\pm$42& 5\% \\
 
& & & & \\

N$_{H}$$^{\ddagger}$ &31.7$\pm$9.1 & 3.66$\pm$2.34 & 61.3$\pm$18.0&$\leq$0.1\% \\

& & & & \\

\hline
\hline
\end{supertabular}

$^{\dagger}$ in units of eV; $^{\ddagger}$ in units of 10$^{22}$ cm$^{-2}$

\normalsize

\vspace{0.3cm}

As previously said, the UM for AGN (Antonucci 1993) predicts that  $\Gamma$, R,
and Ec are observables independent from the inclination angle, thus the two 
classes of Seyfert galaxies should display very similar characteristics. This 
is confirmed by the analysis of the present sample. 
In particular, there are no hints that the distributions of photon 
index $\Gamma$ for the two types of Seyfert are drawn from different
parent populations (P$_{null}$$\sim$90\%). 
Moreover, the photon-index peaks, for both classes, 
between 1.8-1.9 well in agreement with the two-phase models for the production 
of the X-ray in Seyfert galaxies that predicts $\Gamma$$\sim$1.5-2.5 
(Haardt \& Maraschi  1991; Haardt 1993; Haardt, Maraschi \& Ghisellini 1997). 
Few objects have extremely flat spectra with $\Gamma$$\leq$1. 
Type II objects that show so hard X-ray spectra are supposed to be 
Compton-thick sources for which, in the 2-10 keV band, only the reflected/flat 
spectrum is observable. This is the case for NGC 2273 which displays the harder
X-ray spectrum.  This source was first  classified as a Compton-thick object 
by Maiolino et al. (1998). The Seyfert 1s with flattest spectra are NGC 4151 
that is known to have a hard spectrum with complex and variable absorption 
(De Rosa et al. 2006) and Mrk 231 (classified as a type I AGN by Farrah et al. 2003). The latter source shows a very hard X-ray spectrum with 
$\Gamma$$\sim$0.7. This source is also classified as a BAL QSO 
(Smith et al. 1995). A recent spectral analysis in the X-ray
band of this source was presented in Braito et al. (2004). By combining
$XMM$-$Newton$ and $BeppoSAX$ data, these authors speculated that 
the spectrum of the source below $\sim$10 keV is reflection dominated, thus
presenting a case that can be hardly reconciled with the UM of AGN. Moreover, 
Mrk 231 is an ultra-luminous infra-red galaxy. These sources display strong 
starburst activity that can dominate the total X-ray luminosity 
of the galaxy (see Franceschini et al. 2003 and  Ptak et al. 2003 for details 
on this topic) although in Mrk 231 at least 60\% of the observed 
0.5-10 keV flux seems to be due to the AGN component (Braito et al. 2004). 

Less conclusive results are obtained for R and Ec. In both cases, the 
probability of false rejection of the null hypothesis (the two 
distributions are drawn from same parent populations, 
in accordance with  the UM predictions), is P$_{null}$$\sim$5\%. It 
is worth noting, however, that both type I and II objects lay in the same 
range of R and Ec. In particular, the distribution of R for type I objects is 
dominated by a peak of detections between R$\sim$1-1.2 due to the contribution 
of a few sources observed several times (e.g. NGC 5548, IC 4329a). This has 
probably introduced a systematic effect not smeared-out by the relatively 
small number of useful observations (i.e, the ones for which R and Ec have 
been estimated). On the other hand, the result on Ec is most probably 
polluted by the large number of lower limits for Seyfert 2. In fact 
if the same test is performed only on the detections it is obtained that 
P$_{null}$$\sim$35\%, making impossible to assess that the 
two samples are drawn from two different populations.

\begin{figure*}
\centering
\includegraphics[width=5cm,height=3cm,angle=0]{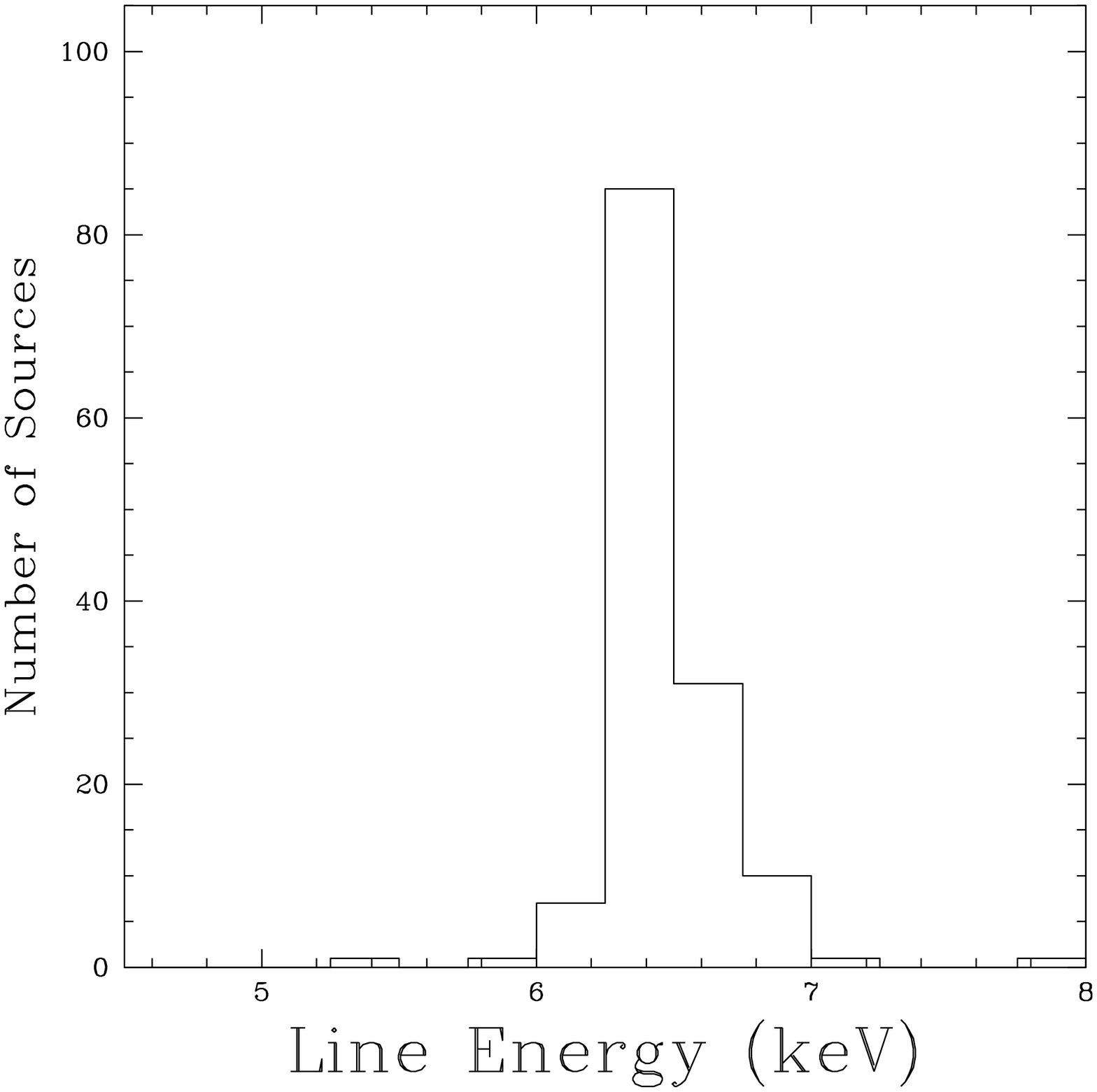}\includegraphics[width=5cm,height=3cm,angle=0]{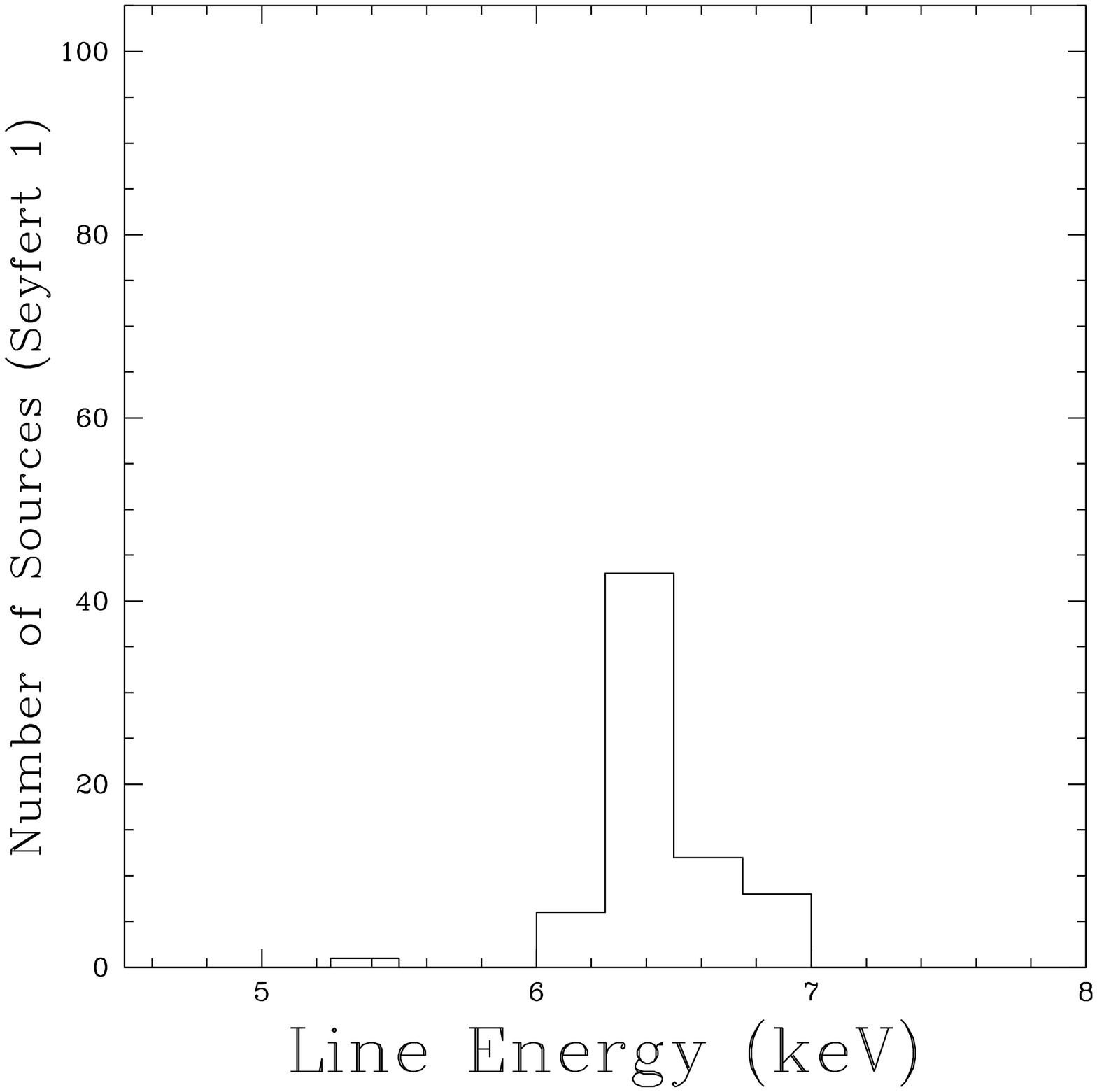}\includegraphics[width=5cm,height=3cm,angle=0]{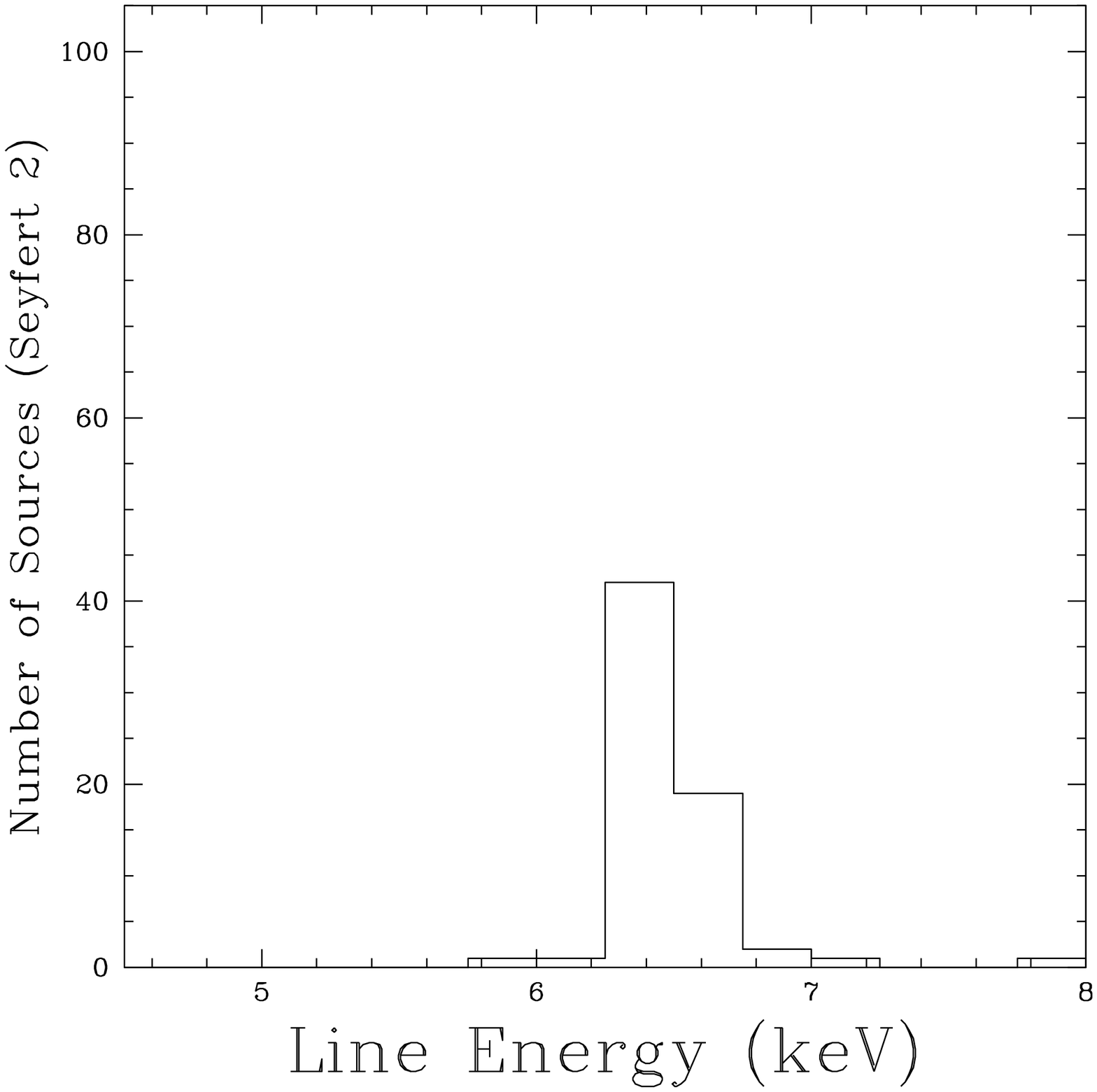}

\includegraphics[width=5cm,height=3cm,angle=0]{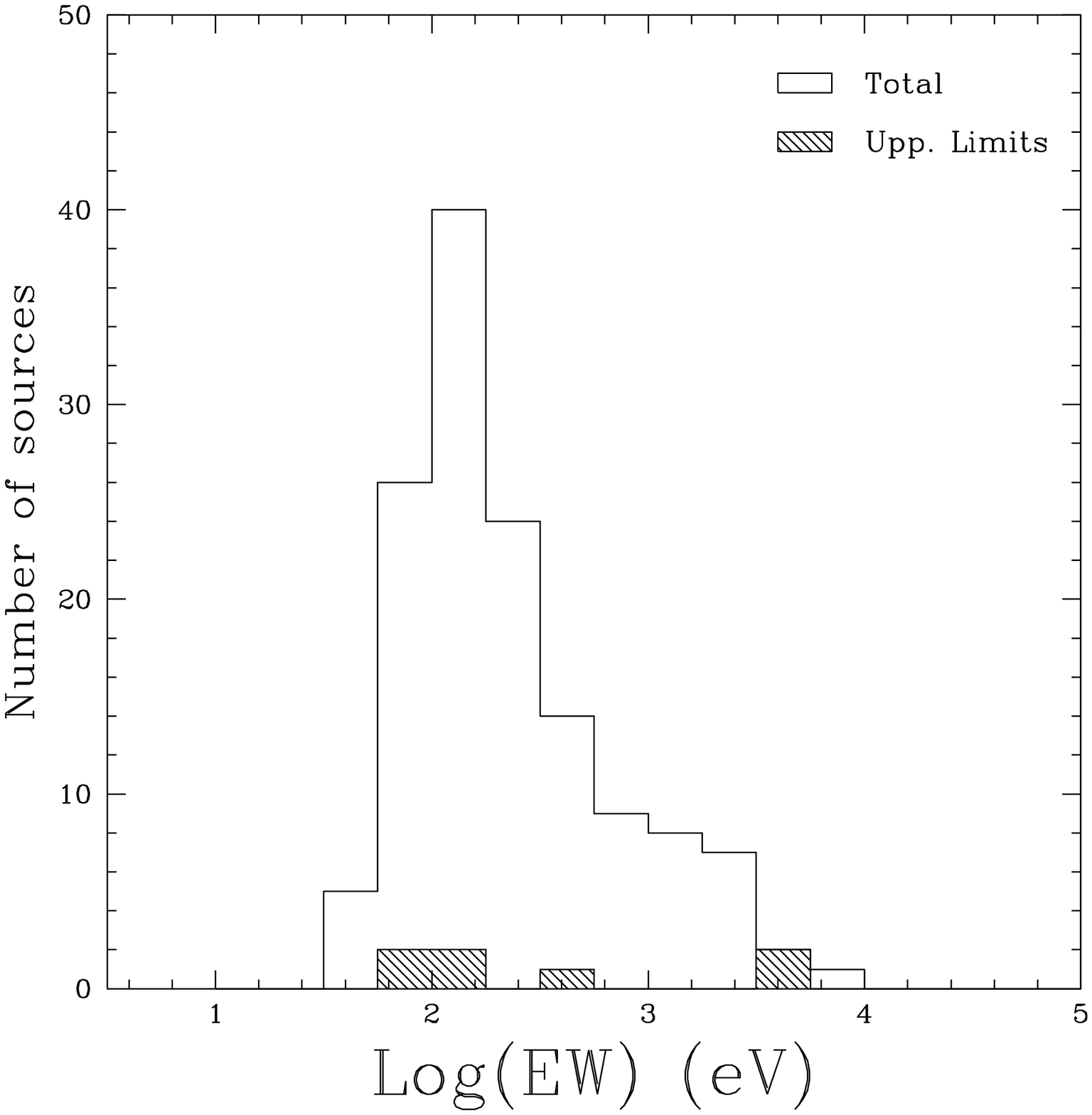}
\includegraphics[width=5cm,height=3cm,angle=0]{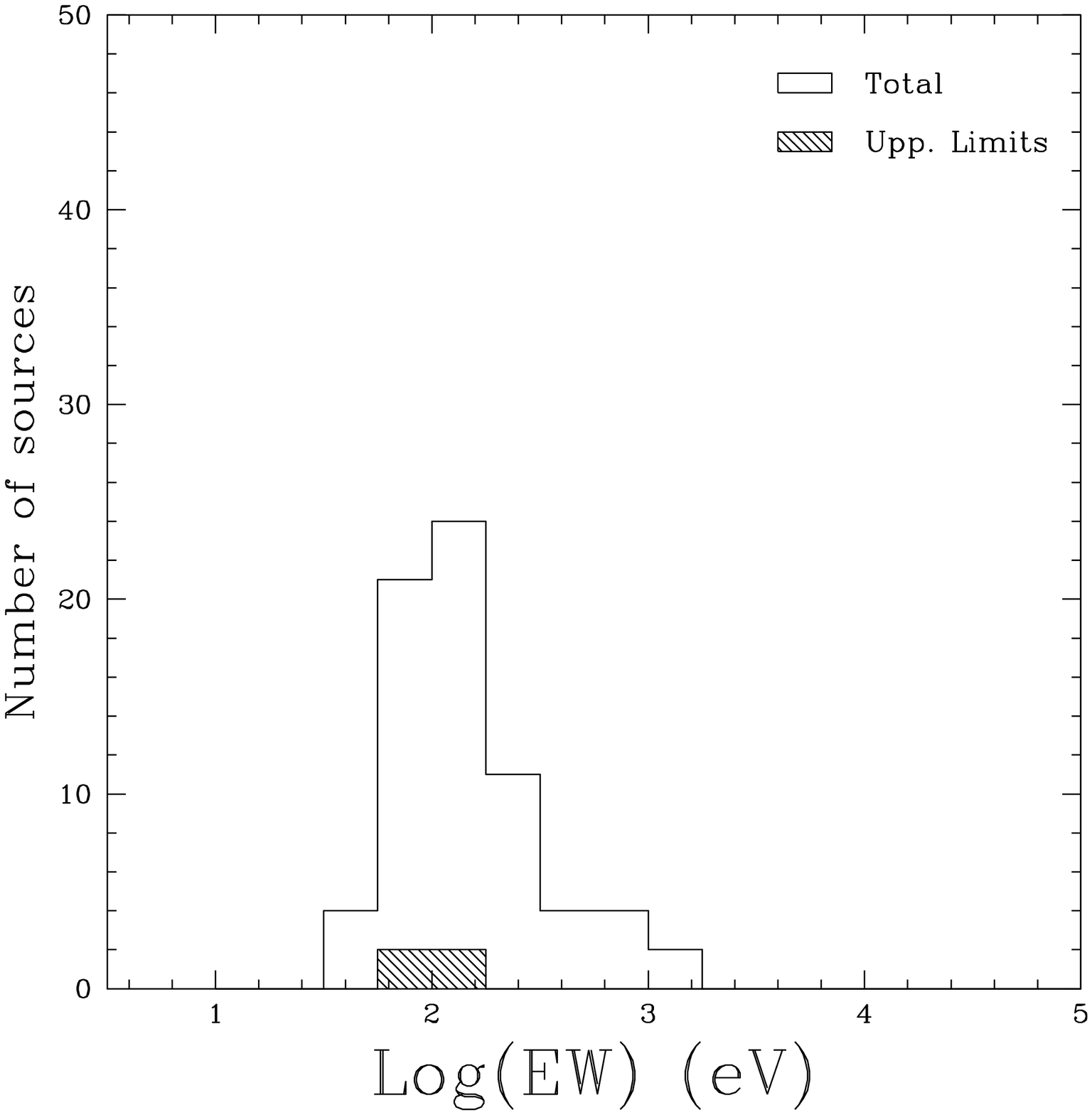}
\includegraphics[width=5cm,height=3cm,angle=0]{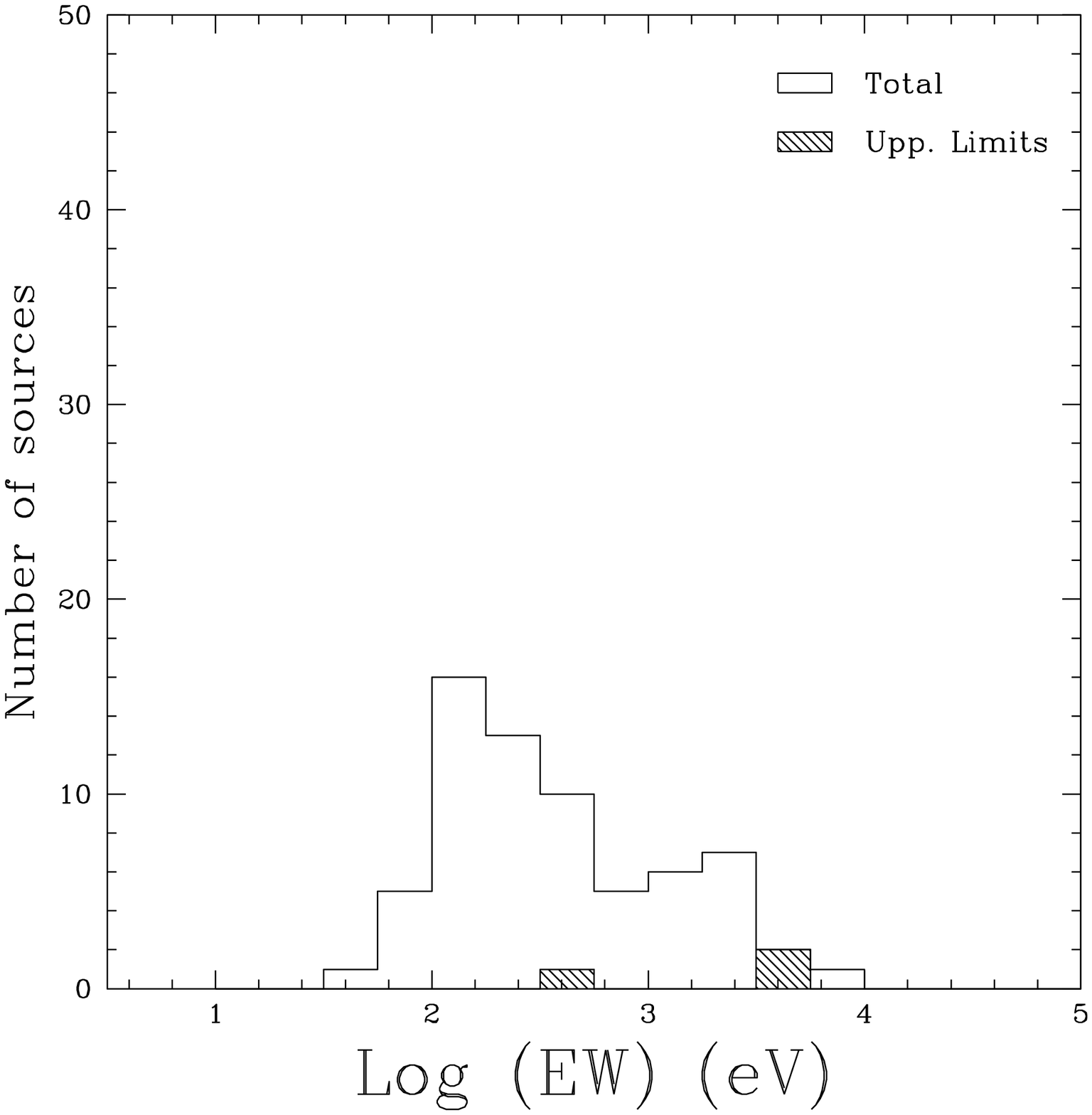}
\caption{Distribution of the Fek$\alpha$ emission line energy centroid ({\it first row}) and of its EW ({\it second row}). Distributions for the whole sample of observations ({\it left panel}), for the type I objects ({\it center panel}) and for type II objects ({\it right panel}).}
\end{figure*}

The result is unambiguously in accordance with UM predictions 
for the absorbing column. In this
case the statistical tests confirm that Seyfert 1 and Seyfert 2 display
very different absorption characteristics with the type II objects being
more heavily  absorbed than Seyfert 1. Again, it is noticeable that a few 
type I objects show high column densities (up to $\sim$10$^{24}$ cm$^{-2}$) and
some Seyfert 2 have low columns (down to 10$^{21}$ cm$^{-2}$). These are not 
new results: the high column of NGC 4151 ($\sim$5$\times$10$^{22}$, De Rosa et al. 2006) is well known.  The highest column
measured in a Seyfert 1 is detected during the June 9, 1998 observation of 
NGC 4051. During this observation the source appeared ``switched-off'' and only a pure reflection component was measured (Guainazzi et al. 1998). The spectral 
fit in Dadina (2007) degenerated between two-solutions, one in which the source was purely reflection-dominated (R$\geq$7) and a second one in which a direct 
component was visible but highly absorbed. The latter scenario was slightly 
preferred by a pure statistical point of view when the 2-200 keV band is
condidered, and, for homogeneity, entered in 
the catalog (Dadina 2007). Nonetheless, when the entire $BeppoSAX$ band 
(0.1-200 keV) is considered, the reflection scheme is preferred (Guainazzi et
al. 1998).

Finally, a number of objects show
upper-limits of the order of $\sim$10$^{22}$ 
cm$^{-2}$ to the absorbing column. This is a selection effect induced by the energy band 
considered using only MECS and PDS data ($\sim$2-100 keV, Dadina 2007). The 
low-energy 
cut-off due to such a column (10$^{22}$ cm$^{-2}$) peaks in fact at E$\sim$2 keV
and only for those objects with a good statistics it is possible
to infer upper limits on the N$_{H}$ below 10$^{22}$ cm$^{-2}$. This is most 
probably responsible of the high value obtain for the average N$_{H}$ in 
type I objects.

\vspace{-0.1cm}

\section{Probing the origin of the FeK${\alpha}$ emission line}

The FeK$\alpha$ line is produced by reprocessing the primary X-ray emission
in matter surrounding the source of hard photons. In the framework of the UM, the origin
of this component can be placed in a number of region
s such as the accretion 
disk, the dusty torus, and the broad-line regions 
(even if this last hypothesis is disfavored by the recent results 
obtained with the $XMM$-$Newton$ and $Chandra$ observatories and 
presented in Nandra 2006). If the line originates in the disk close to
the SMBH, relativistic effects that broaden the resulting line are expected. For the vast majority of the
sources included in the original sample, only narrow component of such feature 
were detected and only in a few cases broad emission lines (e.g. IC 4329a) or 
relativistically blurred features were detected (for example in MCG-6-30-15). 
Thus, the results presented here are essentially based 
on the measured properties of the narrow features.

\begin{figure*}
\centering
\includegraphics[width=5.8cm,height=6.3cm,angle=0]{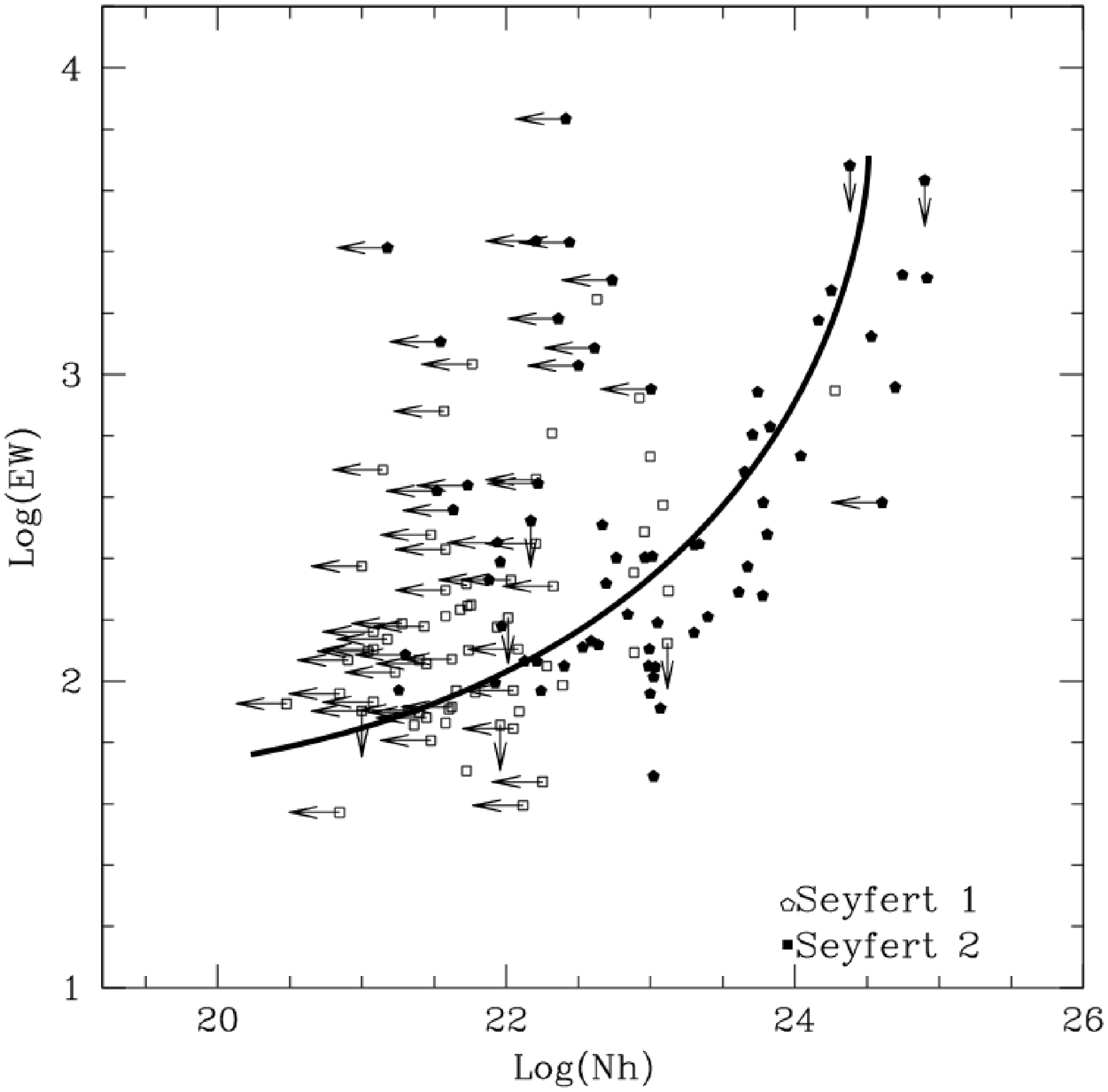}
\includegraphics[width=5.8cm,height=6.7cm,angle=0]{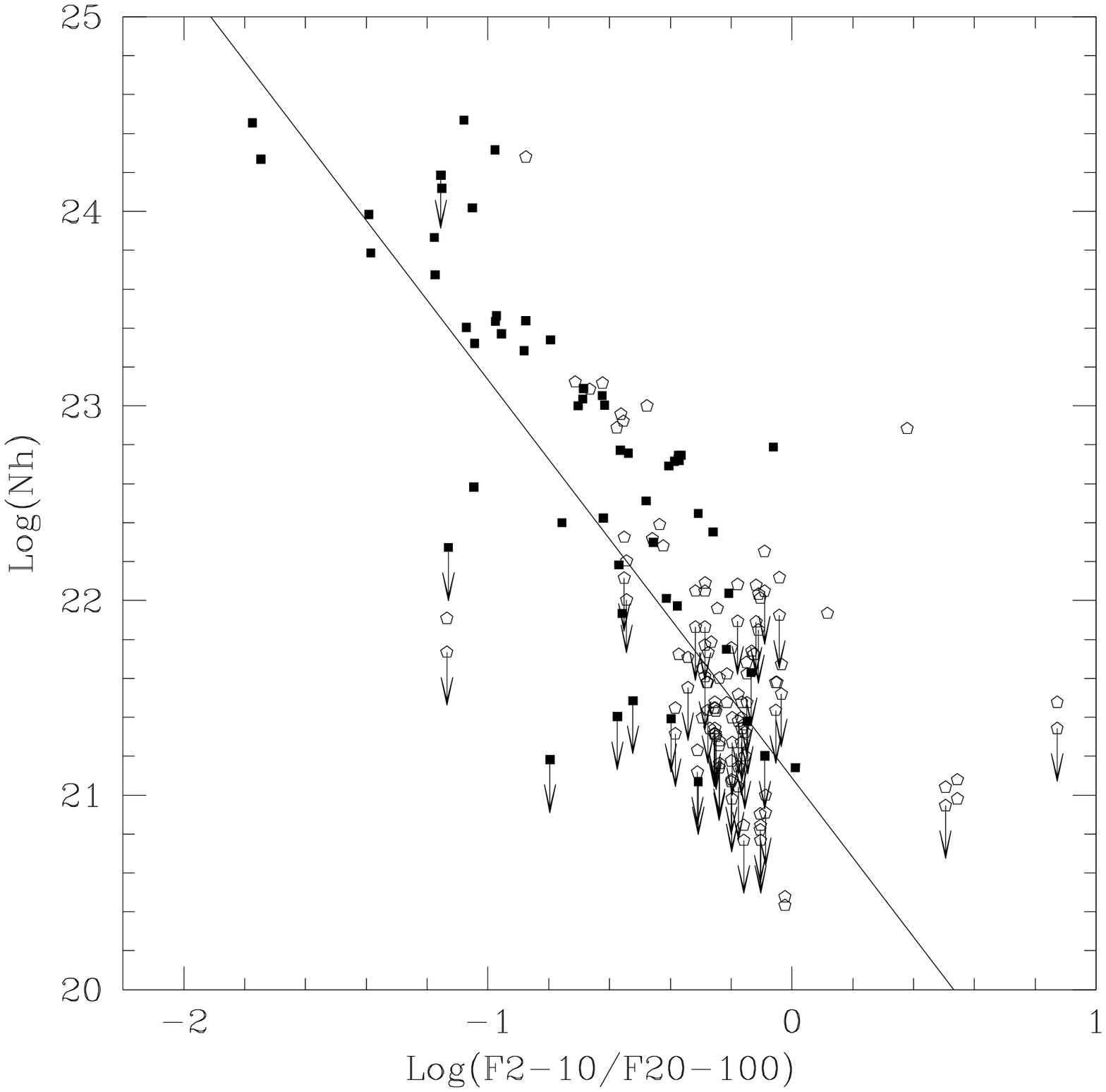}
\includegraphics[width=5.8cm,height=6.7cm,angle=0]{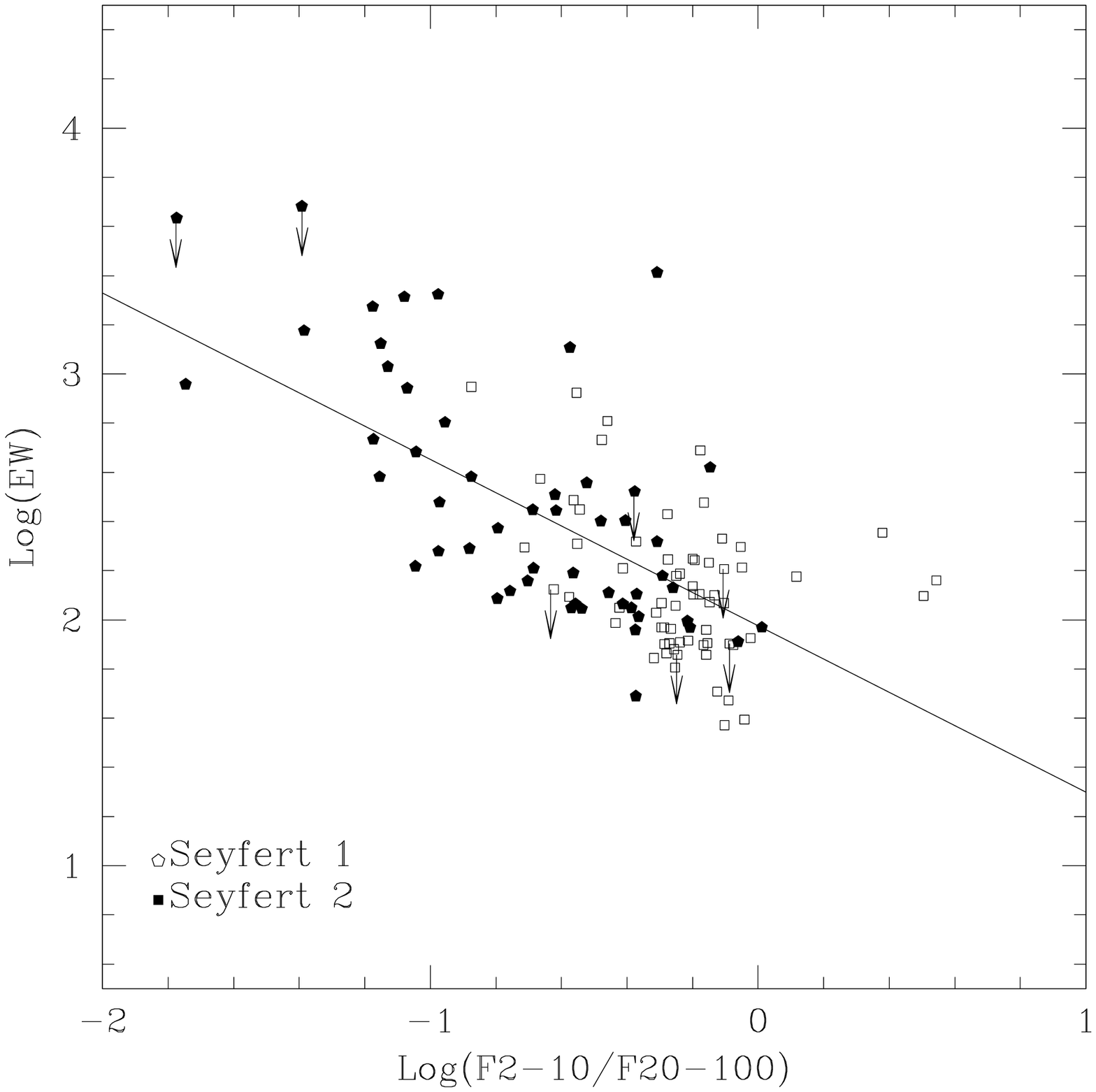}
\caption{{\it Left panel}: Log(EW$_{FeK\alpha}$) vs. Log(N$_{H}$). As expected, the sources are divided in two families: the ones that follow the expectations if the FeK$\alpha$ line is produced in the absorption matter and the candidate Compton-thick ones that display low absorption and large EW. The solid line indicates the prediction by Makishima (1986). {\it Center panel}: Log(F$_{2-10 keV}$/F$_{20-100 keV}$) vs. Log(N$_{H}$). The two quantities are correlated as expected since the 2-10 keV band is strongly affected by the absorption while the 20-100 band is almost free from absorption. The solid line is the best fit obtained with linear regression methods {\it Right Panel}: Log(EW$_{FeK\alpha}$) vs. Log(F$_{2-10 keV}$/F$_{20-100 keV}$). The two quantities are strongly correlated ( P$_{null}$$\leq$0.01). The ratio between X-ray fluxes is a model independent indicator of the absorption affecting the low energy band. Thus, this correlation strongly indicates that the narrow component of FeK$\alpha$ line in emission is indeed produced by the same matter responsible for the absorption. The solid line is the linear regression obtained using Bukley-James method (Isobe et al. 1986).}

\end{figure*}

As shown in figure 2 (first row), the line energy centroid is peaked 
at $\sim$6.4-6.5 keV (see also  table 3) in both type I and II objects. The 
centroid is slightly 
above 6.4 keV but, considering the energy 
resolution of the instrument at these energies 
($\sim$200 eV FWHM, Boella et al. 1997), the results obtained here 
are in agreement  with the line being mainly produced in cold or 
nearly cold matter (ionization state below FeXVII), i.e. in both type I and II objects, by matter 
in the same physical state. However, a well known difference between the two 
classes of objects is the  EW of the narrow FeK$\alpha$ line in type II 
objects, which shows stronger features than type I objects (see second row of 
figure 2 and table 3, Bassani et al 1999, Risaliti et al. 1999, Cappi et al. 2006 and Panessa 
et al. 2006). As  shown in the 
central panel of figure 2, the Seyfert 1 peak at  
EW$_{FeK\alpha}$$\sim$100-200 eV while 
the type II have a broader distribution with an hard tail that reaches values 
well above 1 keV. Also few Seyfert 1 have large values of EW of the 
FeK$\alpha$ line  (above 300-400 
eV). The large EW tail of the Seyfert 1 distribution is composed mainly of 
objects in which the broad components of the FeK$\alpha$ line are detected such
as in MCG-6-30-15, Mrk 841 and IC 4329a. The Seyfert 1 with largest EW is NGC 
4051 during the June 9, 1998 observation when its spectrum was due to pure 
reflection (Guainazzi et al. 1998).

The larger FeK$\alpha$ EW in Seyfert 2 galaxies is in agreement with the 
UM (Antonucci 1993). If the origin of this component is indeed located 
in the dusty torus, than the line EW has to be correlated with 
the  absorber column density. This is indeed what is observed also in this 
sample (see figure 3, left panel). Moreover, the Spearman $\rho$ and 
Kendall's $\tau$ tests indicate that a correlation between the FeK$\alpha$ EW 
and the N$_{H}$ is highly probable for type II objects (P$_{null}$$\leq$0.1\%), 
i.e. for that sources for which we can have direct evidences of the torus 
absorbing column.
 The robustness of the N$_{H}$ 
estimates have been  tested by correlating it with the model independent 
indicator offered by the ratio of the observed fluxes at 2-10 and 20-100 keV 
respectively (center panel of figure 3). The two 
quantities are strongly correlated (generalized Spearman $\rho$ and 
Kendal $\tau$ tests give  P$_{null}$$\leq$0.1\%) with only a few exceptions: 
the pure Compton-thick sources which
are located in the diagram below the majority of the sources. 
This effect is expected since the 
absorbing column can affect the X-ray radiation below $\sim$10 keV while at
harder energies the radiation pierces the matter for columns 
$\leq$3-5$\times$10$^{24}$ cm$^{-2}$. Otherwise, the Compton absorption 
dominates and also photons with energy above 10 keV are stopped since the 
Klein-Nishina regime is reached.

\vspace{0.3cm}

\begin{figure*}
\centering
\includegraphics[width=5.8cm,height=6cm,angle=0]{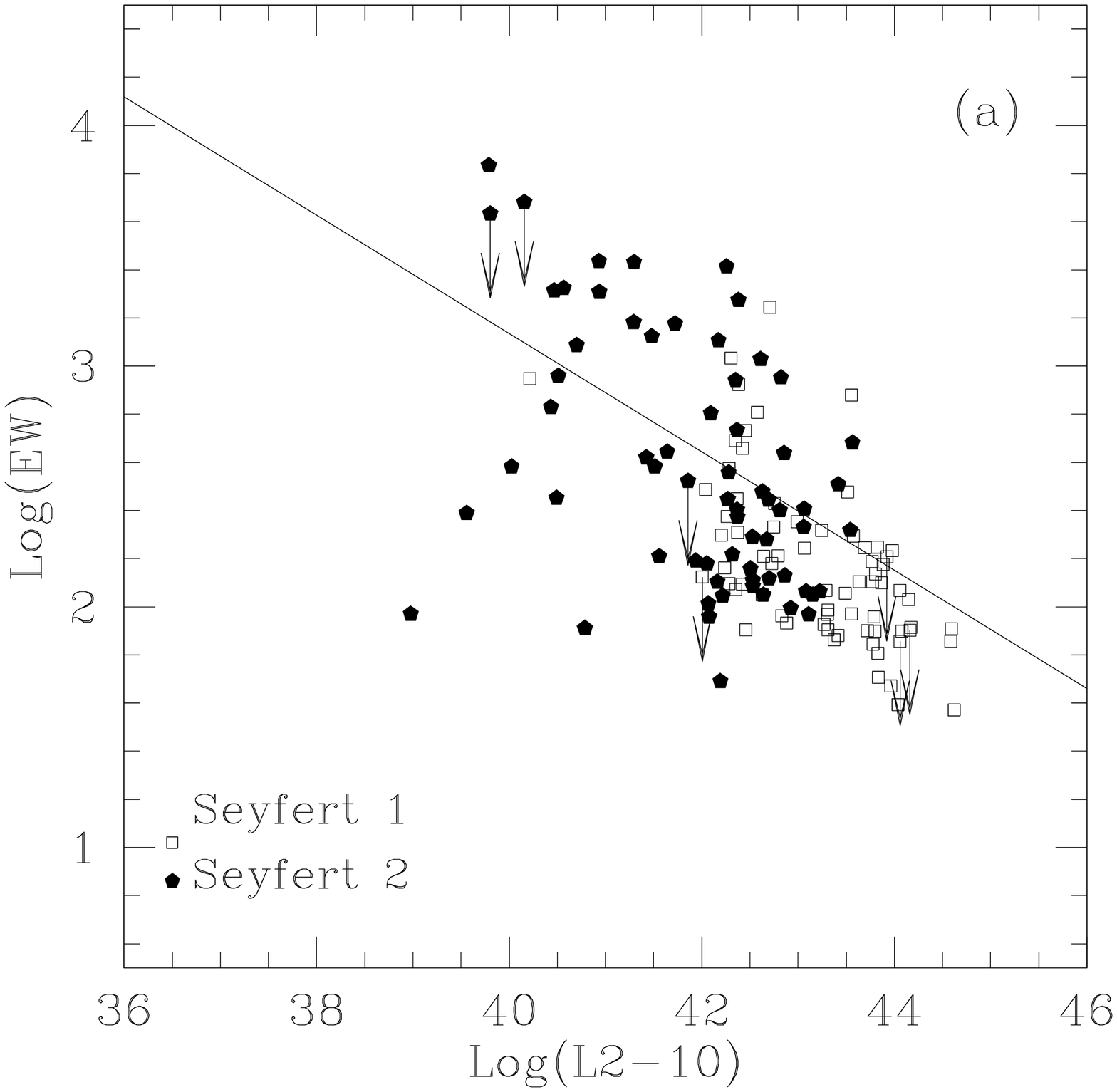}
\includegraphics[width=5.8cm,height=6cm,angle=0]{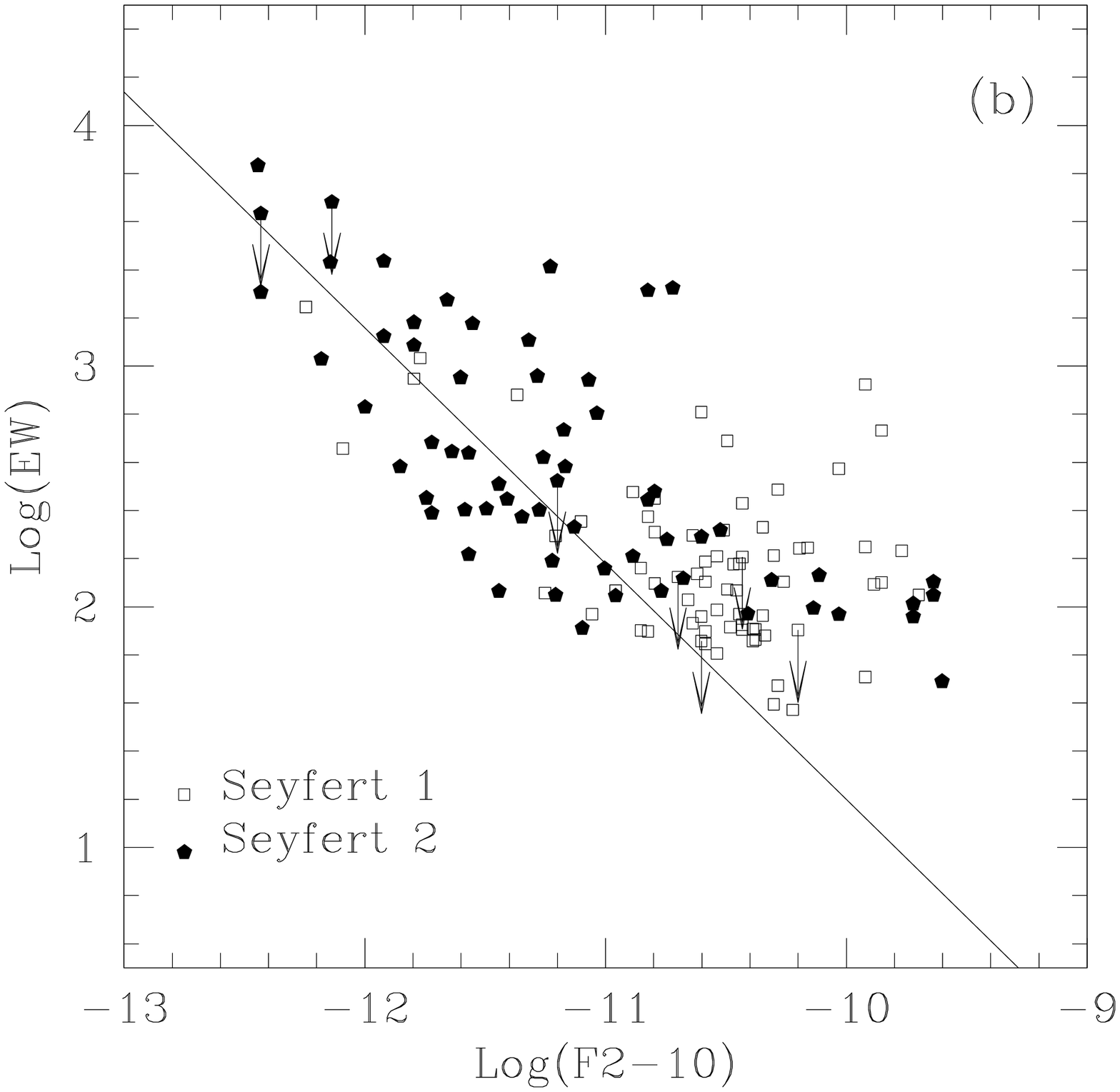}

\includegraphics[width=5.8cm,height=6cm,angle=0]{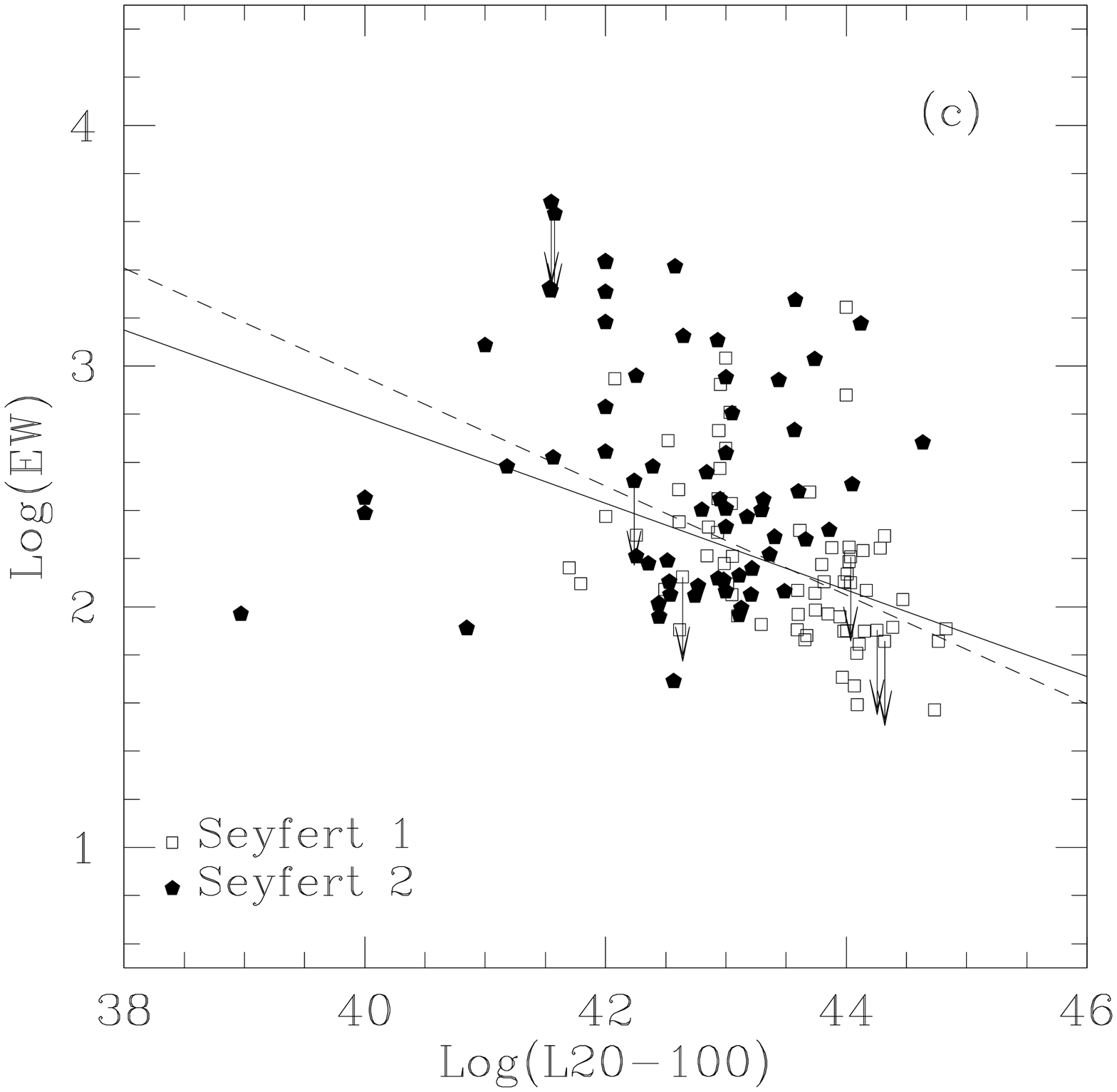}
\includegraphics[width=5.8cm,height=6cm,angle=0]{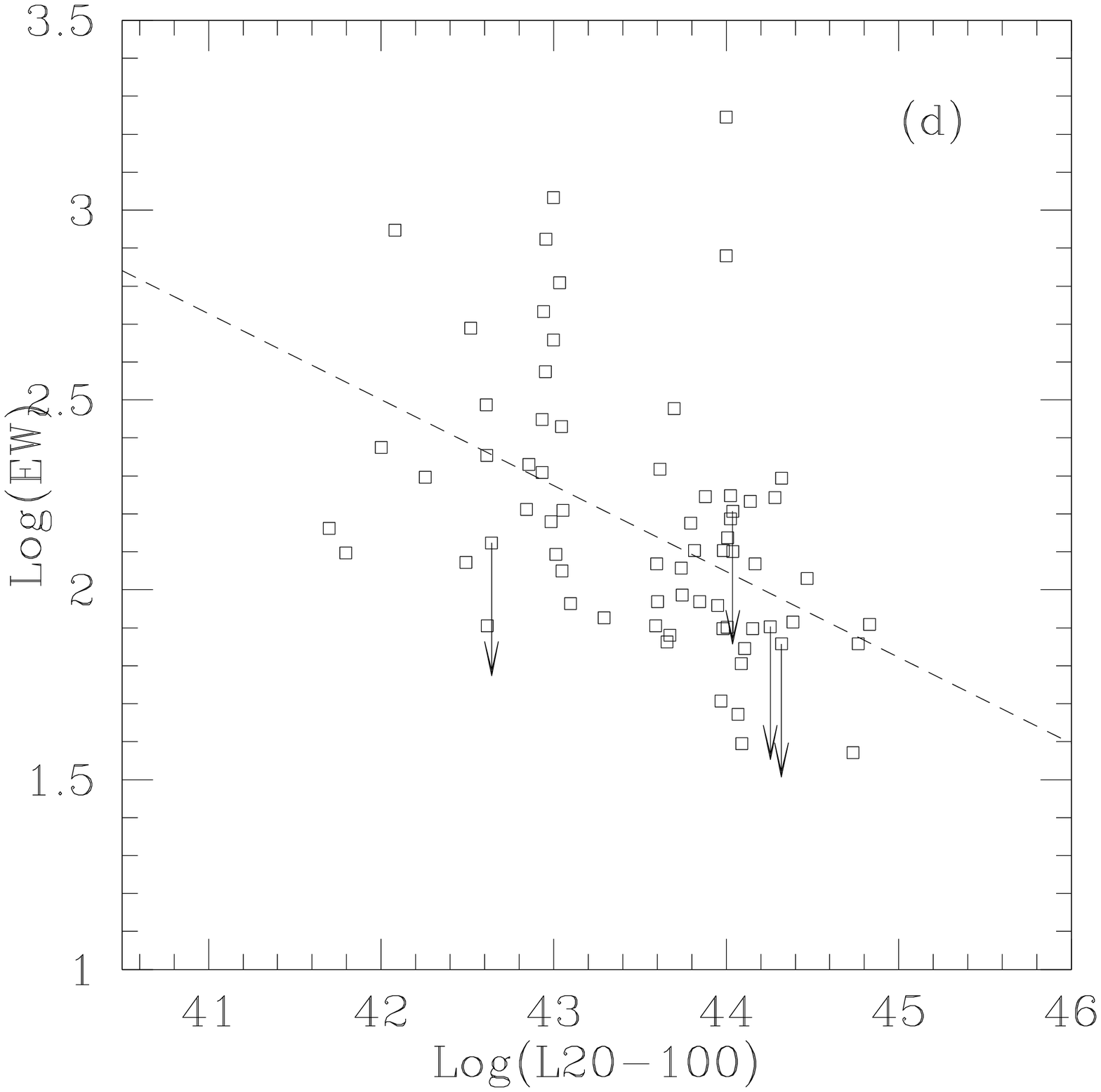}
\includegraphics[width=5.8cm,height=6cm,angle=0]{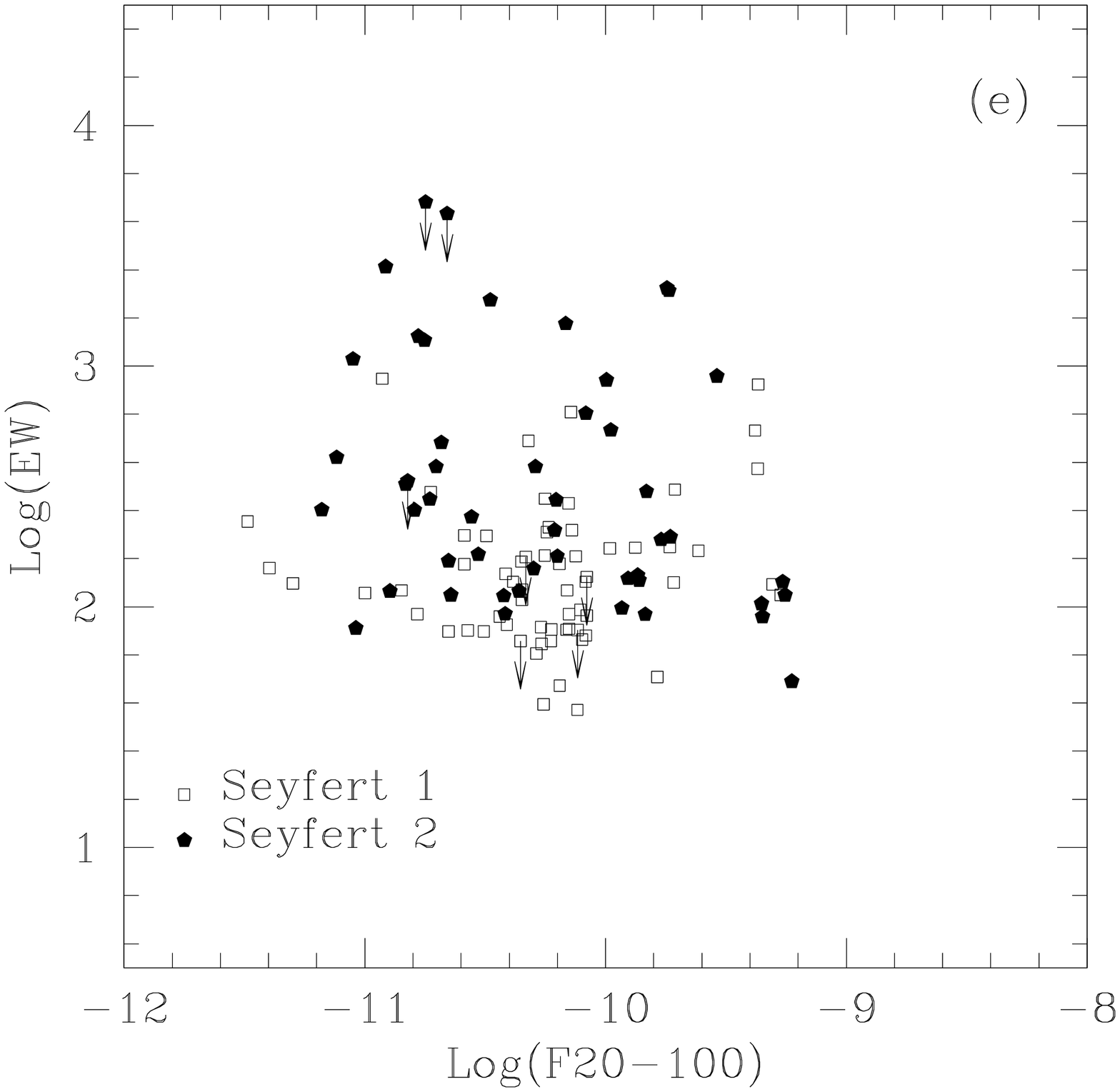}
\caption{{\it Panel (a)}: Log(EW$_{FeK_{\alpha}}$) vs. L$_{2-10, observed}$. {\it Panel (b)}: Log(EW$_{FeK_{\alpha}}$) vs. F$_{2-10, observed}$. {\it Panel (c)}: Log(EW$_{FeK_{\alpha}}$) vs. L$_{20-100, observed}$. {\it Panel (d)}: The same of panel (c) but only for Seyfert 1 objects. {\it Panel (e)}: Log(EW$_{FeK_{\alpha}}$) vs. F$_{20-100, observed}$. Solid lines in panel (a), (b), and (c) are the linear regressions obtained for the whole sample of observations. The dashed line in panels (c) and (e) is the linear regression obtained considering only type I objects. As visible in panel (b) the less scattered relation is obtained considering the 2-10 keV observed flux. The relation is linear, as expected if the 
correlation is due to selection effects, i.e.considering that in faint objects 
only large EW were detectable by the MECS instruments on-board $BeppoSAX$.}
\end{figure*}

\vspace{0.9cm}

\tablecaption{Mean properties of the FeK$\alpha$ emission line. Col. I: spectral parameter; Col. II mean value for the whole sample; Col. III: mean value for Seyfert 1; Col. IV: mean value for Seyfert 2. Col. V: Probability that the parameters of type I and type II objects are drawn from the same parent population.
\vspace{-0.3cm}}
\small
\begin{supertabular}{ l c c c c }
\hline
\hline
& & & & \\

Parameter &Tot. & Seyfert 1$^{\dagger}$ & Seyfert 2$^{\dagger}$ & P$_{null}$ \\

& & & & \\

\hline

& & &  & \\

E$_{FeK\alpha}$$^{\dagger}$ &6.49$\pm$0.02& 6.46$\pm$0.03 & 6.51$\pm$0.03 & 32\% \\ 

& & & & \\

EW$_{FeK\alpha}$$^{\ddagger}$ & 448$\pm$67 & 222$\pm$33 & 693$\pm$195 &  $\leq$1\% \\ 

& & & & \\

\hline
\hline
\end{supertabular}

$^{\dagger}$ in units of keV; $^{\ddagger}$ in units of eV

\normalsize

\vspace{0.2cm}

As stated above, when the EW of the FeK$\alpha$ emission line is tested 
against the measured  N$_{H}$ (left panel of Figure 3) a result in good 
agreement with what predicted by theoretical models is obtained 
(Makishima 1986; Leahy \& Creighton 1993). The majority of the sources, 
in fact, behave as expected if the line is produced by the absorbing matter 
that depress the direct continuum (Makishima 1986).
All the known Compton-Thick sources are located at low  N$_{H}$ and high EW,  in accordance with previous results (Bassani et al. 1999, Risaliti et al. 1999 ). As an additional test, the EW 
of the FeK$\alpha$ line has been plotted 
versus the F$_{2-10 keV}$/F$_{20-100 keV}$ ratio. As expected (see right panel of figure 3), a good correlation ( P$_{null}$$\leq$0.1\% according to generalized Spearman $\rho$ and Kendall $\tau$ tests) is obtained.

\vspace{0.5cm}

These results thus confirm that the properties of the FeK$\alpha$ line
agree with the expectations of the UM for AGN. 
Nonetheless, this is not the only information we have about the Iron line. 
In recent papers (Iwasawa \& Taniguchi 1993; Page et al. 2004, Grandi et al. 
2006; Bianchi et al. 2007) it has been claimed that a X-ray ``Baldwin effect'' (or Iwasawa-Taniguchi effect) is present in AGN when the FeK$\alpha$ intensity 
is probed against the 2-10 keV luminosity. Here this effect is tested 
considering for the first time both the 2-10 keV and the 20-100 keV 
luminosities.

A strong correlation ( P$_{null}$$\leq$0.1\% using Spearman $\rho$ 
and generalized Kendal $\tau$ test) is found when the EW of the FeK$\alpha$ 
line is plotted both against the observed 2-10 keV (panel (a) of figure 4) and 20-100 keV (panel (c) of figure 4) keV luminosities. 
The nature of these correlations, however, is not straightforward, especially 
for the 2-10 keV luminosity. In this energy band the effect of the absorber is 
very important. As previously demonstrated, the EW of the Iron line correlates
with N$_{H}$, but stronger absorbing columns imply lower fluxes. 
Moreover, when the relation between the FeK$\alpha$ line EW
and the 2-10 keV flux is investigated (panel (b) of figure 4), it is found 
that the two quantities are correlated (P$_{null}$$\leq$1\%).
This is explainable only in terms of instrumental effects, since the 
sensitivity of the instruments to narrow features decreases with the source's
flux. Thus, to be detected in faint objects, the 
FeK$\alpha$ line must be strong enough. Since the original sample
is limited to the local Universe, this correlation in flux acts, at least 
partially, also in the EW vs L$_{2-10 keV}$ relation.

\begin{figure*}
\centering
\includegraphics[width=5.8cm,height=6cm,angle=0]{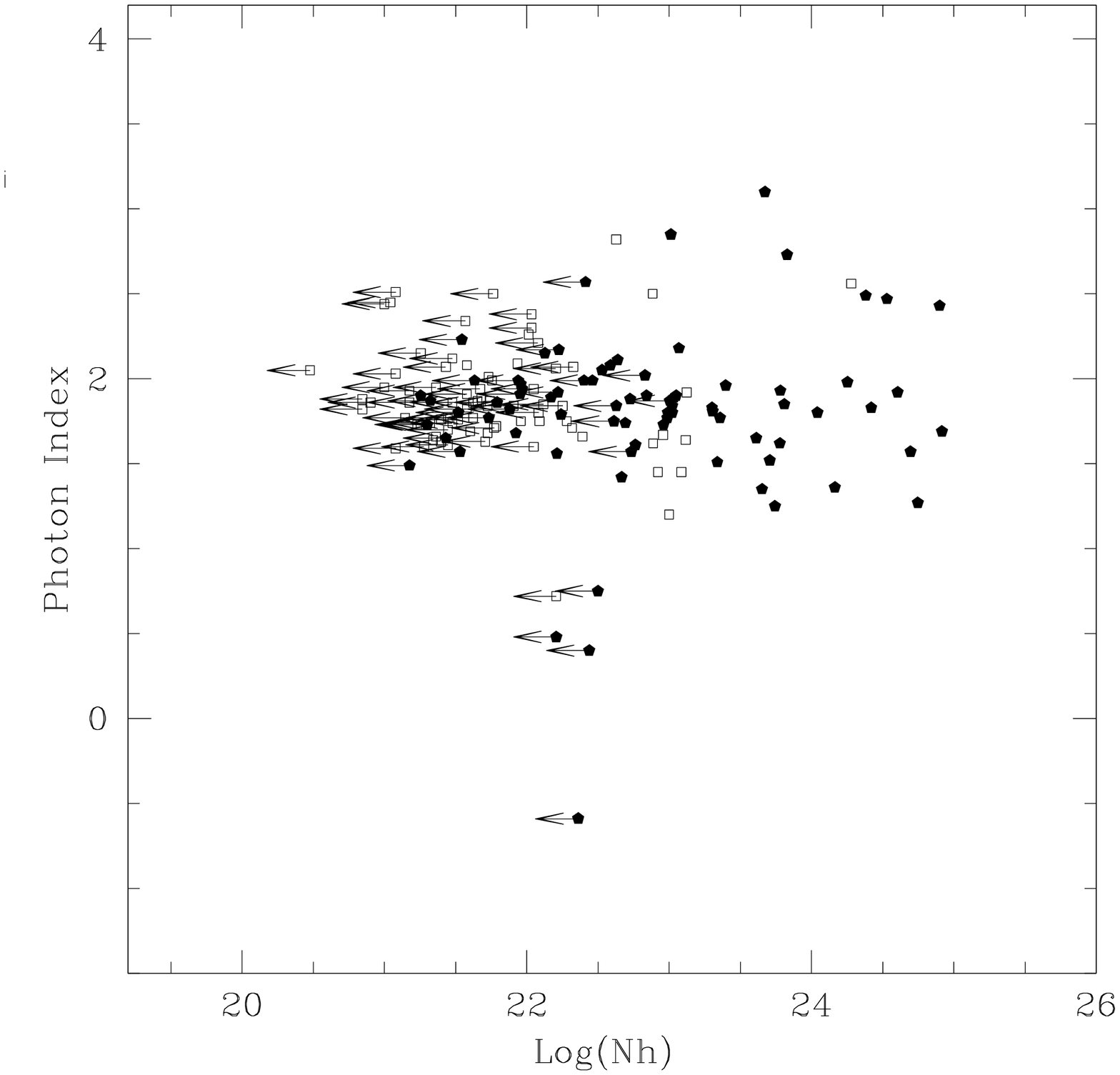}
\includegraphics[width=5.8cm,height=6cm,angle=0]{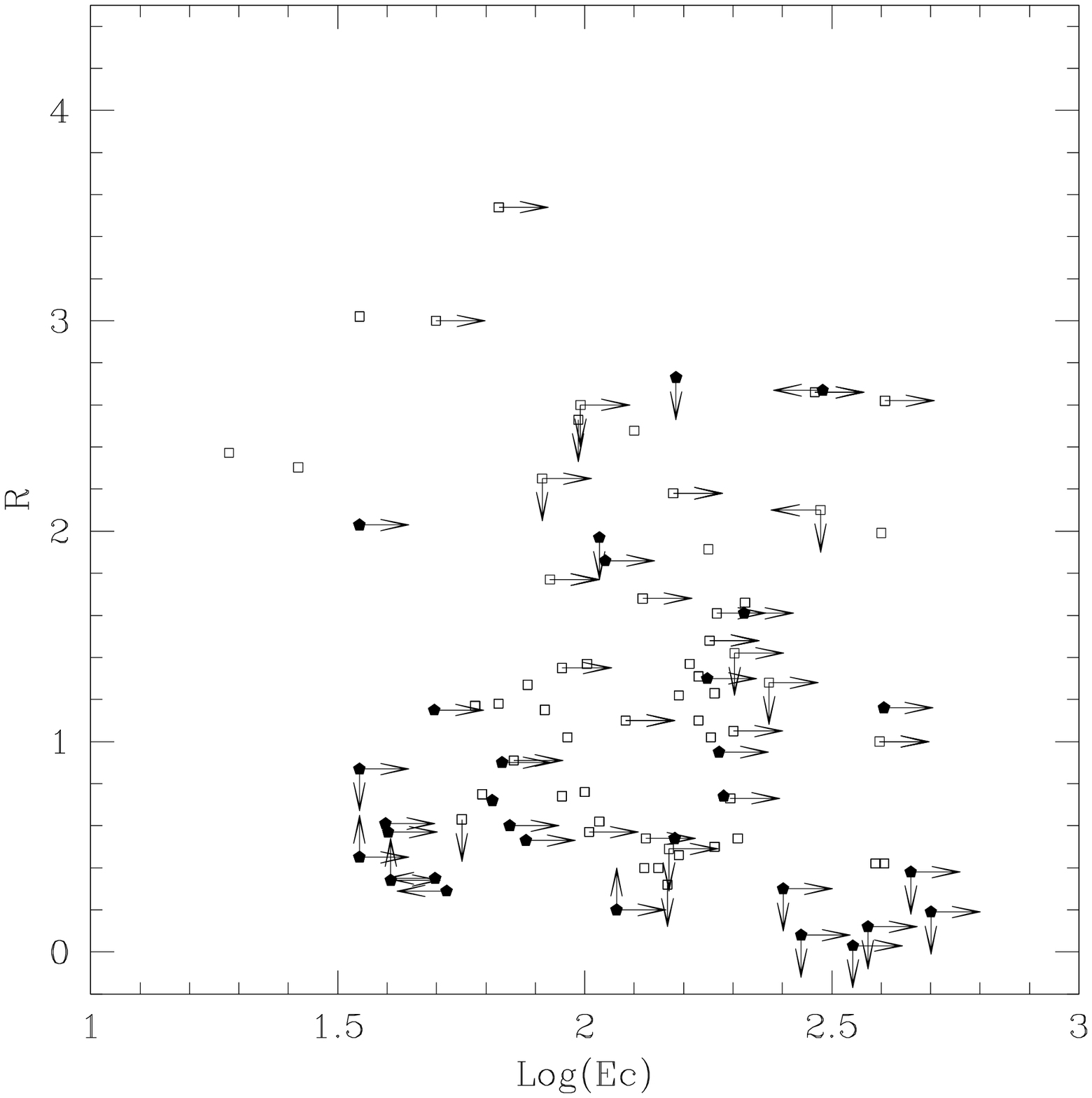}
\includegraphics[width=5.8cm,height=6cm,angle=0]{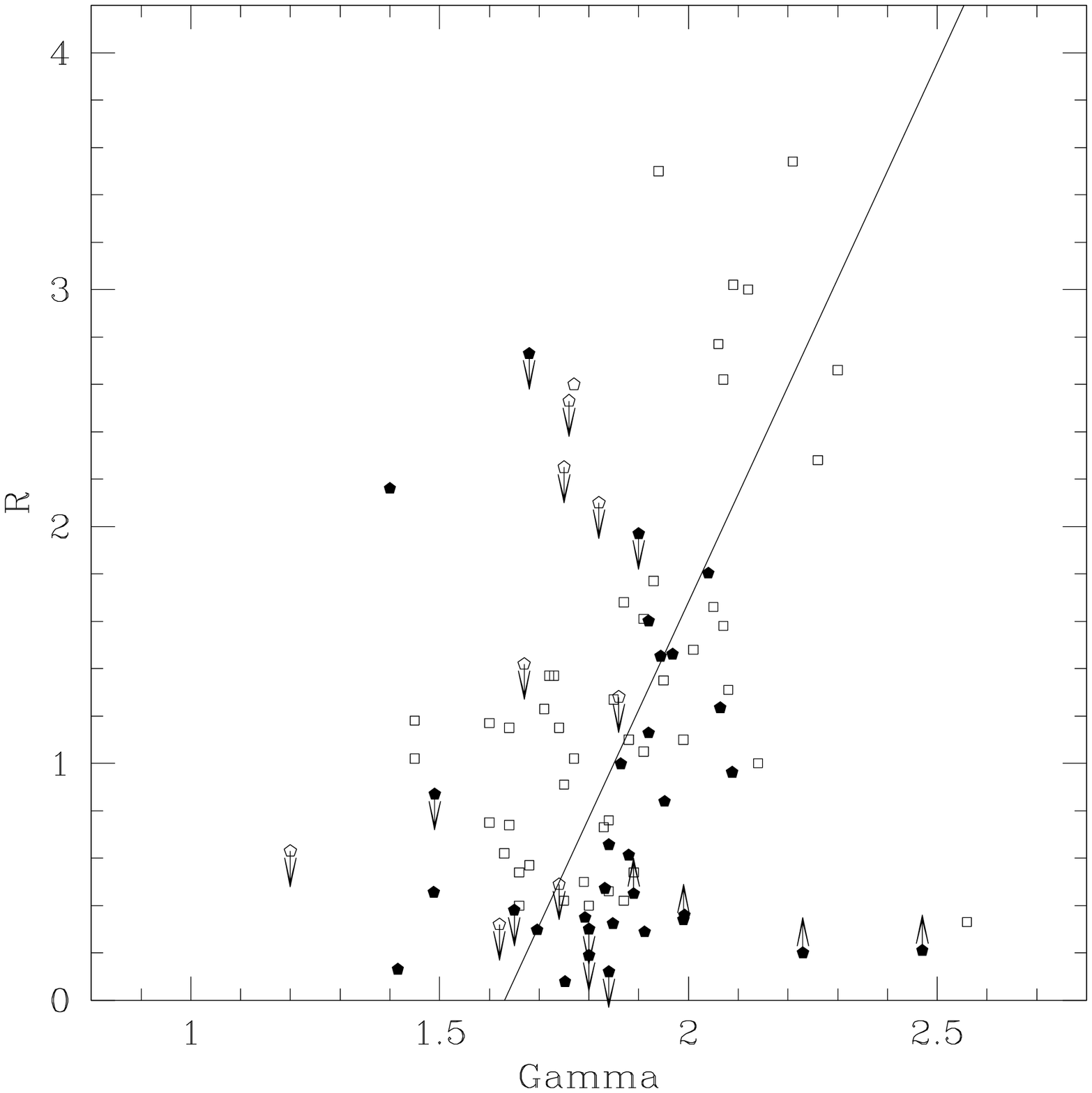}
\caption{{\it Left panel:} Photon index $\Gamma$ plotted versus the measured N$_{H}$ (in units of cm$^{2}$). No significance trend relating these two quantities is found; {\it center panel:} R vs. Ec (in units keV), no correlation is found between these quantities; {\it right panel} R vs. $\Gamma$. The two quantities are correlated with a high significance level (P$_{null}$$\leq$1\%). The line is the linear regression best fit.}
\end{figure*}

This N$_{H}$ effect should be negligible when the 20-100 keV band is 
considered.
In fact, in this case, one expects to find a correlation between the observed 
20-100 keV luminosity and the EW of the FeK$\alpha$ line only in the most 
extreme cases, i.e. for the ``pure Compton thick'' objects. Apart from 
NGC 1068, these sources are too weak to be detected by the PDS, thus 
unable to drive the relation observed in plot (c) of figure 4. 
Moreover, panel (e) of figure 4 indicates that the EW of the iron line is not
related to the 20-100 keV flux thus excluding that the ``Baldwin effect'' 
measured using the 20-100 keV band is due to instrumental selection effect as 
it happens for the F$_{2-10 keV}$. 
On the other hand, in the 20-100 keV band the reflection-hump 
at $\sim$30-40
keV contributes to the observed flux. If the origin of the FeK$\alpha$ line is 
due to the same 
matter responsible of the reflection, one should expect that the EW of the 
Iron line should increases as the reflection component augments the 20-100 keV
flux (i.e. the Iron line EW should correlate with the 20-100 keV flux). 
Thus, the net contribution of the reflection component should act in the 
opposite direction to that observed (i.e. larger iron line EW at smaller 
20-100 keV flux). If the two classes of Seyfert galaxies are analyzed 
separately it is obtained that a strong correlation is found for Seyfert 1 
(P$_{null}$$\leq$0.1\% using both Spearman $\rho$ and generalized Kendal 
$\tau$ tests, panel (d) of figure 4) while no correlation
is evident for type II objects (P$_{null}$$\sim$80\%). 
This result is not unexpected  since the EW of the obscured sources are
boosted by the suppression of part of the underlying continuum. To conclude, 
the presence of a X-ray Baldwin effect for Seyfert 1 is unambiguously 
confirmed by present data if the L$_{20-100 keV}$ is  considered and it has 
the following relation:

\small
\vspace{0.4cm}

\hspace{-0.4cm }Log(EW)=-(0.22$\pm$0.05)$\times$Log(L$_{20-100}$)+11.91$\pm$2.52\hspace{0.6cm}(1)

\vspace{0.4cm}
\normalsize

The slope of the relation found in this work 
is in agreement with what previously obtained by Page et al. (2004) (EW$\propto$(L$_{2-10}$)$^{-0.17\pm0.08}$) 
using a sample containing both radio-quiet and radio-loud objects.  Present 
result is also consistent with what found by Zhou \& Wang (2005) (who used a sample containing both radio-quiet and radio-loud objects founding 
 EW$\propto$(L$_{2-10}$)$^{-0.15\pm0.05}$) and 
Bianchi et al. (2007) (who used only radio-quiet objects obtaining EW$\propto$(L$_{2-10}$)$^{0.17\pm0.03}$). On the contrary, Jiang, Wang \& Wang (2006) found that, excluding the radio-loud AGN from a sample similar to the one used by Page et al. (2004), 
found that EW$\propto$(L$_{2-10}$)$^{0.10\pm0.05}$. It is worth recalling here, that the the present sample is composed by both radio-loud and radio quiet sources (Dadina 2007). In particular, it contains 7 radio-loud Seyfert 1, and only for 5 of them the 20-100 and iron line data are available. Nonetheless, 
the presence of these sources in the sample does not affect the FeK$\alpha$ EW vs. L$_{20-100}$ relation (EW$\propto$(L$_{2-10}$)$^{0.21\pm0.05}$ excluding radio-loud type I objects).

The  origin of the X-ray ``Baldwin effect'' is unclear. In the light bending 
scenario (Miniutti \& Fabian 2004) the height of the source above the
accretion disk determines the degree of beaming along the equatorial plane
of the high energy emission. Because of relativistic effects, the closer the 
source is to the disk, the greater will be the fraction of  
X-rays beamed in the equatorial plane (i.e. towards the disk) and 
correspondingly lower will be the observed flux.
Thus, the EW of the relativistically blurred FeK$\alpha$ line produced
in the inner regions of the disk and the EW of the narrow Iron line produced
in the outer parts of the disk would appear enhanced in low-state sources.  

On the contrary, Page et al. (2004) 
speculated that this effect could indicate that luminous 
sources are surrounded by dusty tori with lower covering factor thus 
pointing towards a torus origin of most of the narrow FeK$\alpha$ line. 
The present work supports this view. The FeK$\alpha$ line EW of the Iron line
correlates with the observed N$_{H}$ as predicted by theory (Makishima 1988;
Lehay \& Creighton 1993; Ghisellini, Haardt \& Matt 1994). 
Moreover, also the case of the extremely low state of NGC 4051 
(Guainazzi et al. 1998) included in this dataset seems to point in this 
direction. The huge EW of the narrow FeK$\alpha$ line recorded in this 
observation, in fact, is typical of Compton-thick type II objects, but the 
line does not show  evidence of relativistic 
broadening due to the contribution of the inner orbits of the accretion disk.

\section{The $\Gamma$-R relation.}

In the fitting procedure some parameters may degenerate given the 
interdependence among them. This is the case, for example of the photon index 
with the column density for low statistics observations. The same 
could happen for the determination of R and Ec, since R introduces in the 
spectrum a bump peaked between 20-40 keV and declining at higher energies 
where the Ec may appear. 
To check if the results presented here are affected by such effects, 
the correlations between these parameters have been studied and the results 
are presented in Figure 5.

Left  panel of Figure 5 shows how, on average, the estimate 
of $\Gamma$ is not affected by the simultaneous determination of the absorbing 
column. In fact, no trend is observed between $\Gamma$ and N$_{H}$. Obviously, 
this does not imply that this is true for each single source included in the 
original sample. On the other hand, this is an expected result since 
the broad band of $BeppoSAX$ should reduce this spurious effect.  
Similar results are obtained also when the $\Gamma$ vs. Ec, and R vs. 
Ec (center panel of Figure 5) relations are 
investigated. In these cases the Spearman $\rho$ and 
Kendall's $\tau$ tests do not sustain the existence of 
a relation between these quantities (P$_{null}$$\sim$15\%).
All these indications suggest that, if any, the possible 
degeneracy in the fitting procedure did not introduce strong 
spurious relations between the spectral parameters.

However a strong correlation  (P$_{null}$$\leq$0.01\%) is recorded 
between $\Gamma$ and R that are linked by the following relation:

\small
\vspace{0.4cm}

R=(4.54$\pm$1.15)$\times$$\Gamma$-(7.41$\pm$4.51) \hspace{3.2cm} (2)

\vspace{0.4cm}
\normalsize

It is hard to define if this correlation is the result
of a systematic effect or not since it is possible that these two quantities 
degenerate in the fitting procedure. Flat power-law with 
small reflection could be described also by steep power-law plus strong 
reflection.  The total absence of similar correlations between
$\Gamma$ and Ec and R vs. Ec seems to suggest that this correlation 
is indeed real.

A similar relation was previously found using 
$Ginga$ and $RXTE$ data (Zdziarski, Lubinski, \& Smith 1999; Gilfanov, 
Churazov \& Revnivtsev 1999). Zdziarski, Lubinski, \& Smith (1999) 
interpreted it as evidence of thermal Comptonization as origin of
X-rays providing 
that the optical-UV seed photons were mainly produced by the same material 
responsible for the reflected component. In this case, in fact, the cooling 
rate of the hot corona is directly linked to the power-law slope. But the 
cooling rate is also related to the angle subtended by the reflector. This 
result is also in agreement with predictions of models that consider mildly 
relativistic outflows driven by magnetic flares (Beloborodov 1999). 
More in general, Merloni et al. (2006) demonstrated that any geometry in which 
the hot, X-ray emitting 
plasma, is photon starved (i.e. geometries of the accretting systems  in
which the accretion disk is only partially covered by the Comptonizing 
plasma such as patchy coronae, inner ADAF plus outer disks etc.) will produce 
hard X-ray spectra, little soft thermal emission and 
weak reflection component. On the other hand, geometry corresponding to a very 
large covering fraction of the cold phase, have strong soft emission, 
softer spectra and strong reflection fraction (Collin-Souffrin et al. 1996). 
Thus, moving along the $\Gamma$ vs. R relation it implies moving from lower to
higher accretting systems.

\section{Summary and conclusions}

The average properties of Seyfert galaxies in the local
Universe (z$\leq$0.1) as seen by $BeppoSAX$  has been investigated, analyzing 
the sample of
objects presented in Dadina (2007). Multiple observations of single objects 
were treated independently, i.e. the multiple measurements of parameters were not
averaged for statistical purposes. This method has been chosen since, in the 
framework of the simplest version of UM for AGN (Antonucci, 1993) the 
AGN are thought to be very similar to each-other and only 
orientation/absorption effects and the activity-level of the targets could 
introduce observational 
differences between different objects. 
In this scenario, the monitoring of a single source could be reproduced by the 
observations of many sources in different states and ``vice versa''.

$BeppoSAX$ offered the advantage of a useful X-ray broad energy band. Data 
studied here fall in  the 2-100 keV band for a majority of objects. 
This advantage has been used to investigate the properties
of the high-energy continuum of Seyfert galaxies. 
As stated in the 
previous paper (Dadina, 2007), the basic template was  
a power-law with a high energy cut-off plus a reflection component (namely
PEXRAV model in XSPEC). The results of this analysis can be 
summarized as follow:

\begin{itemize}

\item the average slope of the power-law is 1.84$\pm$0.03 for the entire sample
of objects. Considering the two families of AGN separately, it turns-out that $\Gamma$=1.89$\pm$0.03 for type I objects (including Seyfert 1, 1.2 and 1.5) 
while $\Gamma$=1.80$\pm$0.05 for type II objects (including Seyfert 1.8, 1.9 
and 2);

\item the average value of the relative reflected-to-direct normalization
parameter R is 1.01$\pm$0.09 with a slight difference between the two classes
of Seyfert galaxies (R=1.23$\pm$0.11 for type I objects and R=0.87$\pm$0.14
for the type II ones); 

\item the high energy cut-off was measured to be Ec=287$\pm$24 (Ec=230$\pm$22 keV for Seyfert 1 and Ec=376$\pm$42 keV for Seyferts 2);

\item as expected and as known from previous works, the absorbing column is 
very different in the two classes of objects. On average N$_{H}$$\sim$3.66$\times$10$^{22}$ cm$^{-2}$  type I and N$_{H}$$\sim$6.13$\times$10$^{23}$ cm$^{-2}$ for type II AGN. The high mean value obtained for Seyfert 1 is caused by a selection effect induced by the energy coverage of the MECS+PDS instruments (2-100 keV).

\item evidence of a X-ray Baldwin effect is found in Seyfert 1 galaxies when 
the EW of the  FeK$\alpha$ line is plotted against the 20-100 keV 
luminosity. 

\item a significant correlation has been found between R and $\Gamma$.

\end{itemize}

These results are well in agreement with the basic assumptions of the 
UM for AGN (Antonucci 1993). In fact, no differences are measured in the 
observables that are supposed to be isotropic while the absorbing column 
seems to be the only discriminator between the different types of Seyfert 
galaxies.
This reflects also in the properties of the FeK$\alpha$ line. No difference
is measured in the line centroid (see table 3) between the two classes of 
Seyfert galaxies.
Type II objects, however, display more intense features 
(EW=222$\pm$33 eV for Seyfert 1 and EW=693$\pm$195 eV for Seyfert 2).
The  physical origin of the X-ray ``Baldwin effect'' here measured for Seyfert 1 using 
the 20-100 luminosity is unclear. Both light bending 
(Miniutti \& Fabian  2004) and torus models (Page et al. 2004) are consistent  
with present data even if the strong relation of the FeK$\alpha$ line EW 
in type II objects with the absorption column indicates
that the most of the narrow line component should be due to the torus.
Finally, the measured $\Gamma$-R relationship is consistent with thermal 
Comptonization models.

\begin{acknowledgements}

I thank G.G.C. Palumbo and M. Cappi for helpful discussion and for 
careful reading of the previous versions of the manuscript. I also thank the 
ASDC people for their wonderful work in mantaining the $BeppoSAX$
database. I really thank the referee for her/his helpful comments and 
suggestions that contributed to improve the quality of the manuscript.
Financial support from ASI is aknowledged.

\end{acknowledgements}

\end{document}